\documentclass[aps,reprint,onecolumn,floatfix,nofootinbib,12pt]{revtex4-2}
\usepackage{hyperref}
\hypersetup{
  breaklinks=true, % splits links across lines
  colorlinks=true, % displays links as colored text instead of blocks
  pdfusetitle=true, % \title and \author values into pdf metadata
}
\usepackage{float}
\usepackage{epsfig}
\usepackage{graphicx}
\usepackage{amsmath}
\usepackage{amsfonts}
\usepackage{amssymb}
\usepackage{bm}
\usepackage{bbold}
\usepackage{epsfig}
\usepackage{graphicx}
\usepackage{color}
\usepackage{pictex}
\usepackage{mathtools}
\usepackage{extarrows}
\usepackage{footnote}
\usepackage{footmisc}
\usepackage{tabularx,colortbl}
\usepackage{subcaption}
\usepackage{setspace}
\begin{document}
\title{Hierarchy of percolation patterns in a kinetic replication model }
%:  neural network application}

% Author Orchid ID: enter ID or remove command
%\newcommand{\orcidauthorA}{0000-0002-1745-2447} % Add \orcidA{} behind the author's name
%\newcommand{\orcidauthorB}{0000-0000-0000-000X} % Add \orcidB{} behind the author's name

\date{\today}

\author{Pavel Ovchinnikov}
%\email{ovchinnikov.pa@dvfu.ru}
\affiliation{Institute of High Technologies and Advanced Materials, Far Eastern Federal University, Russky Island, Ajax 10, Vladivostok 690922, Russia}
\affiliation{Institute of Applied Mathematics, Far Eastern Branch, Russian Academy of Science, Radio 7, Vladivostok 690041, Russia}
\affiliation{Bogoliubov Laboratory of Theoretical Physics, Joint Institute for Nuclear Research, Dubna 141980, Russia}

\author{Konstantin Soldatov}
\affiliation{Institute of High Technologies and Advanced Materials, Far Eastern Federal University, Russky Island, Ajax 10, Vladivostok 690922, Russia}
\affiliation{Institute of Applied Mathematics, Far Eastern Branch, Russian Academy of Science, Radio 7, Vladivostok 690041, Russia}
\affiliation{Bogoliubov Laboratory of Theoretical Physics, Joint Institute for Nuclear Research, Dubna 141980, Russia}

\author{Vitalii Kapitan}
\affiliation{Department of Statistics and Data Science, National University of Singapore, 21 Lower Kent Ridge Road, 119077, Singapore}

\author{Gennady Y. Chitov}
\email{Corresponding author: gennady.chitov@gmail.com}
\affiliation{Bogoliubov Laboratory of Theoretical Physics, Joint Institute for Nuclear Research, Dubna 141980, Russia}
\affiliation{D\'{e}partement de physique, Universit\'{e} de Sherbrooke, Sherbrooke, Qu\'{e}bec, J1K 2R1 Canada}

%
%xxxxxxxxxxxxxxxxxxxxxxxxxxxxxxxxxxxxxxxxxxxxxxxxxxxxxxxxxxxxxxxxxxxxxxxxxxxxxx
%
\begin{abstract}
The model of a one-dimensional kinetic contact process with parallel update is
studied by the Monte Carlo simulations and finite-size scaling. The goal was to reveal the structure of the hidden percolative patterns (order parameters) in the active phase and the nature of transitions those patterns emerge through. Our results corroborate the earlier conjecture that in general the active (percolating) phases possess the hierarchical structure (tower of percolation patterns), where more complicated patterns emerge on the top of coexistent patterns of lesser complexity. Plethora of different patterns emerge via cascades of continuous transitions. We detect five phases with distinct patterns of percolation within the active phase of the model.
All transitions on the phase diagram belong to the directed percolation universality class, as confirmed by the scaling analysis.  To accommodate the case of multiple percolating phases the extension of the Janssen-Grassberger conjecture is proposed.
\end{abstract}
\maketitle

%
%
%xxxxxxxxxxxxxxxxxxxxxxxxxxxxxxxxxxxxxxxxxxxxxxxxxxxxxxxxxxxxxxxxxxxxxxxxxxxxxx
%xxxxxxxxxxxxxxxxxxxxxxxxxxxxxxxxxxxxxxxxxxxxxxxxxxxxxxxxxxxxxxxxxxxxxxxxxxxxxx
%%%%%%%%%%%%%%%%%%%%%%%%%%%%%%%%%%%%%%%%%%%%%%%%%%%%%%%%%%%%%%%%%%%%%%%%%%%%%%
%
\section{Introduction}\label{Intro}
%xxxxxxxxxxxxxxxxxxxxxxxxxxxxxxxxxxxxxxxxxxxxxxxxxxxxxxxxxxxxxxxxxxxxxxxxxxxxxx
%xxxxxxxxxxxxxxxxxxxxxxxxxxxxxxxxxxxxxxxxxxxxxxxxxxxxxxxxxxxxxxxxxxxxxxxxxxxxxx
%
%
Percolation is an example of phase transition involving nonlocal order which probes connectivity of a system \cite{Stauffer:1992}.
The evolution of many kinetic models (kinetic contact process with parallel update,
probabilistic cellular automata, PCA)  in $1+1$ (space-time) dimensions results in 2D directed percolative landscapes. The applications of such models range from statistical physics, critical phenomena, condensed matter to biology, ecology, quantitative finance or network theory \cite{Hinrichsen:2000,Hinrichsen:2006,Grassberger:2015,Newman:2010,Barabasi:2016,Wetterich:2021,Stepinski:2023}.

It has been shown  \cite{Chitov:2015,Chitov:2016}  that the active (percolating) phases of several $(1+1)$ kinetic processes and some other 2D models of percolation
possess numerous hidden geometric orders characterized by distinct  order parameters (percolative patterns), emerging at specific critical points of continuous phase transitions. The order parameters are the capacities of corresponding percolating clusters spanning through whole system in the time direction (or, equivalently, the density of the connected active sites in  the stable state of directed processes). These transitions belong to the directed percolation (DP) universality class,
and they are confirmed unambiguously  \cite{Chitov:2016} even for the well-known DP and contact process models which have been studied for about six decades or so \cite{Hinrichsen:2000,Hinrichsen:2006}.
To the best of our knowledge, such cascades of transitions with different patterns within the percolating phase have not been reported in the literature,
other than \cite{Chitov:2015,Chitov:2016}. More qualitatively on the results \cite{Chitov:2015,Chitov:2016}: the raw 2D data generated by the probabilistic models reveal distinct types of modularity (patterns) in different regions of the parametric space \textit{within} the active (percolating) phases of those models. (See, e.g., left panels in Figs.~\ref{fig:dipole_pattern_example},\ref{fig:quadrupole_pattern_example},\ref{fig:PL_pattern_example}.) It is possible to detect distinct percolative order
parameters probing those patterns, on the top of the conventional order parameter related to a simply connected cluster of the system's size. Detection of the hidden percolative patterns bears quite close analogy with the search for modules (communities) in networks (graphs) which are by the construction connected (they have a giant component) \cite{Fortunato:2010}.

The multitude of conceivable percolation patterns implies the existence of infinite cascades of geometric transitions in the parametric space. Moreover,
the hierarchical structure (tower of percolation patterns) appears to be a generic feature of percolating phase,  and it should be looked for in other models of statistical mechanics and kinetics. This is a very important conclusion for the whole theory of percolation. More broadly, such hierarchy should exist in a generic distributed system/network and can be used to quantify its connectedness. This hypothesis  has vast implications for different fields, and in particular for networks.

Network science is a very active interdisciplinary field with virtually ubiquitous
applicability. It ranges from telecommunications, electric power supply, Internet,
warehouse logistics, transportation, engineering  to physics, biology, neurology, and sociology \cite{Dorogovtsev:2008,Newman:2010,Sporns:2010,*Sporns:2012,Barthelemy:2011,Barabasi:2016,FornitoBull:2016,Simpson:2023}. A particular important aspect is the proper functioning  and resilience of networks: see  \cite{Newman:2010,Stanley:2010,*Stanley:2012,*Stanley:2013,*Havlin:2021,Barabasi:2016,Dey:2019,Lou:2023,Kotlarz:2024} and more references there. One of the motivations to study the hidden structure of the active phases of $(1+1)$ processes or 2D percolation landscapes, is to reveal a similar hierarchy in generic networks. We will come back to this point in the concluding section.

The notion of the percolative backbone (we will call it ``pattern" in the present study) was defined in \cite{Chitov:2016} as percolation on the backbone cluster spanning through whole system. This cluster consists of renormalized (coarse-grained) nodes made out of a chosen subset of nodes of the original lattice (graph) with a particular choice of filling.\footnote{These ideas are somehow similar to the earlier proposals of high-density \cite{Leath:1978,Turban:1979,Timonin:2018} or $k$-clique percolation
\cite{Derenyi:2005,*Derenyi:2007} on different lattices, graphs or networks. Similar notions of ``motifs" were used in the context of brain connectivity \cite{Sporns:2010,*Sporns:2012} and other complex networks \cite{Dey:2019}.
We should point out that the multiple transitions found in networks or other models of percolation differ from those  reported here and in  \cite{Chitov:2015,Chitov:2016}, since the former are either related to the singularities of a single order parameter or due to coupling of different networks.}
The percolation occurs with respect to the bonds connecting active nodes of the backbone cluster.\footnote{In case of 2D lattice considered in the present work, the backbone cluster can be simply called the renormalized lattice.}
There are two points to be clarified in the above definition:\\
1). The attribution of activity to the renormalized node (filled/empty) allows different choices. For instance, for the case of $2 \times 2$ plaquettes to be identified as new renormalized nodes, one can select a particular half-filled configuration (quadrupole) as an active renormalized node and discard all other fillings as a renormalized empty state; or select a configuration where plaquette is either fully filled or one out of four original sites is empty, see Fig.~\ref{Patterns}. \\
2). The bonds between renormalized sites do not need to coincide with their counterparts in the original lattice. This is plainly obvious in the context of networks: say, on the map of a town one can consider a backbone cluster of acquainted couples.  Or,  for the model considered in this work: its spreading rules allow to admit percolating bonds (due to correlations) between the nearest and next-nearest renormalized nodes.

To build up working algorithms for unfolding the hidden hierarchy of percolative patterns,
we will study the replication model (a version of PCA) proposed in \cite{Chitov:2015}.
The attractive feature of this model is that its active phase demonstrates evolution in the parametric space from quite distinguished antiferromagnetic-like pattern to a fully occupied lattice. This was suggestive  for the search of a new phase  (denoted as  $D^+$ in Fig.~\ref{PDiag}) with the dipole percolating pattern, found earlier in Ref.~\cite{Chitov:2015}. In this paper we report six percolating phases on the phase diagram of the model, see Fig.~\ref{PDiag}, and the hierarchical structure of the active phase,  where more complicated percolating patterns emerge on the top of coexistent patterns of lesser complexity, see Fig.~\ref{OPs}.
The results demonstrate clearly the key role of both elements in the emergence of a percolating pattern: 1) local order of primary sites within a renormalized node; 2) the range of percolation admitted between locally ordered (that is active) nodes of the backbone.

The conjecture \cite{Chitov:2016} that there is potentially an infinite number of patterns hidden in the percolating phase is not rigorously proven yet.\footnote{This is a problem analogous to counting the number of possible partitions of a graph with $N$ vertices into $k$-clusters, $1 \leq k \leq N$, see, e.g. \cite{Fortunato:2010}. It grows faster than the exponential of $N$ at $N \gg 1$.  \label{InfCasc}}
But even if so, construction of an infinite tower of percolation patterns is not very practical or useful for comparative analysis of percolation landscapes or their resilience -- two problems very relevant for applications. The important task is to construct a judicious case-motivated set of percolative patterns to classify a given landscape. The selection of patterns cannot be defined rigorously, this is a judgment call based mainly on the numerical experiments for a concrete model.
For the model studied in this work, we could have searched for the $4 \times 4$ plaquettes or probably even more sophisticated and/or bigger patterns.
We restricted our analysis to the judicious set of quite simple and visually distinguished patterns built on the blocks of sizes $2 \times 1$ and $2 \times 2$.

%
%
%xxxxxxxxxxxxxxxxxxxxxxxxxxxxxxxxxxxxxxxxxxxxxxxxxxxxxxxxxxxxxxxxxxxxxxxxxxxxxx
%xxxxxxxxxxxxxxxxxxxxxxxxxxxxxxxxxxxxxxxxxxxxxxxxxxxxxxxxxxxxxxxxxxxxxxxxxxxxxx
\section{Model and Methods}\label{Methods}
%xxxxxxxxxxxxxxxxxxxxxxxxxxxxxxxxxxxxxxxxxxxxxxxxxxxxxxxxxxxxxxxxxxxxxxxxxxxxxx
%xxxxxxxxxxxxxxxxxxxxxxxxxxxxxxxxxxxxxxxxxxxxxxxxxxxxxxxxxxxxxxxxxxxxxxxxxxxxxx
%
%
In this paper we study a one-dimensional kinetic contact replication process (a version of probabilistic cellular automaton, PCA).
The state of the system at discrete time step $t = 0,1, \ldots, T$ is specified by the occupation numbers $n_{i,t} = 0,1$ (empty/filled), $i = 1,\ldots, N$,
with periodic boundary conditions (PBC) in the spatial direction.
At $t=0$ the initial random configuration of $n$-s is cast.
The Monte Carlo simulations with parallel update (i.e., the configurations at all sites are updated simultaneously in one time step) are used to numerically mimic stochastic evolution of the model.
The latter is governed by the transfer probabilities defined as the probability to have a filled site  $n_{i,t+1}$ at the time $t+1$ given the configuration of neighboring sites $n_{i-1,t}$, $n_{i,t}$, and $n_{i+1,t}$ at the time $t$. We denote it as $P(n_{i,t+1}|n_{i-1,t},n_{i,t},n_{i+1,t})$.
The model's transfer probabilities are:
\begin{eqnarray}
  P(1|1,1,1) &=& P(1|0,1,1)=P(1|1,1,0)=P(1|0,1,0)=p,  \nonumber \\
  P(1|0,0,1) &=& P(1|1,0,0)=q,  \nonumber\\
  P(1|1,0,1) &=& 1-(1-q)^2=q(2-q), \nonumber\\
  P(1|0,0,0) &=& 0,~~ P(0|*,*,*)=1-P(1|*,*,*)~.
\label{ModDef}
\end{eqnarray}

\begin{figure}[h]
\centering{\includegraphics[width=7.2cm]{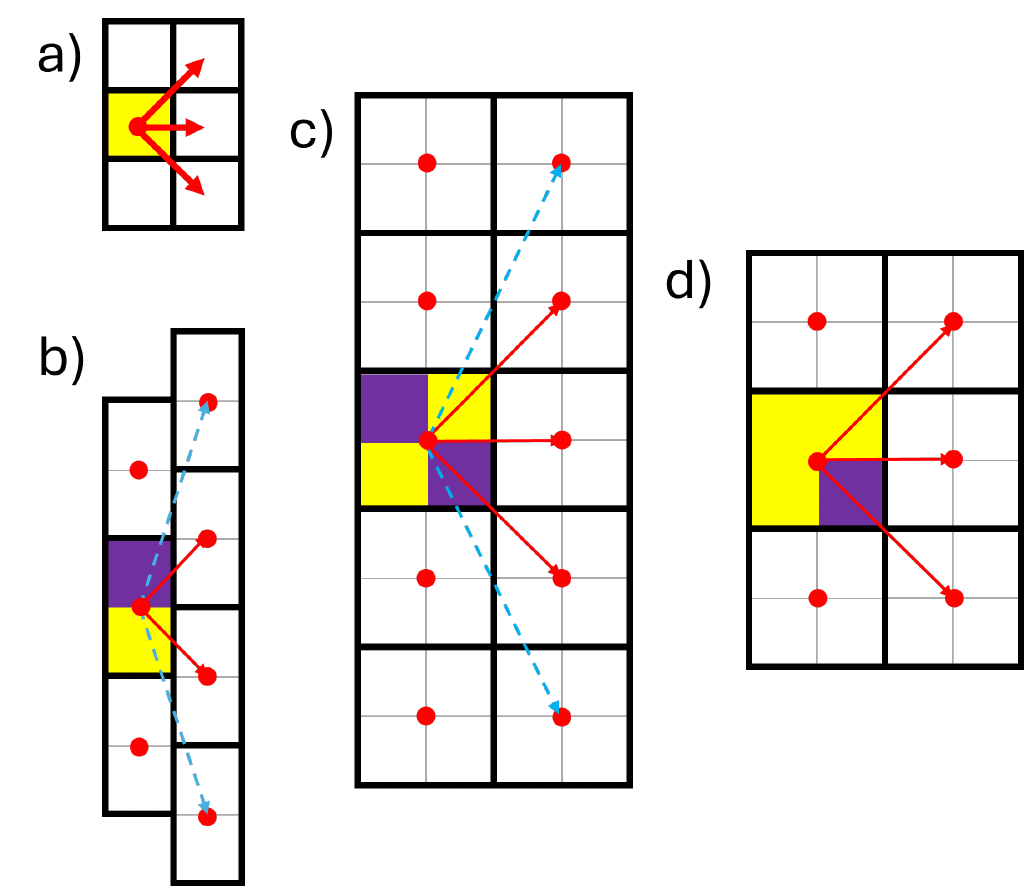}}
\caption{Illustration of different directed percolation patterns considered in this work. Active (filled) sites are shown in yellow; empty sites - violet. a) percolation on the sites of the original lattice occurring along directed bonds shown in solid red. b) percolation between dipoles (bold solid black). Their occupation numbers are defined by Eq.~\eqref{Dns}, nodes of the dipole lattice shown by bold red dots. Percolation pattern in $D$ phase accounts for nn dipoles (solid red lines). In $D^+$ phase the pattern includes nn and nnn (dashed light blue) dipoles. c) Similar for $Q$ and $Q^+$ phases, quadrupole occupation numbers residing on the nodes of plaquette lattice (bold red dots) are defined by Eq.~\eqref{Qns}. $2\times 2$ plaquettes shown in bold solid black. d) $PL$ phase accounts for percolation between nn (solid red lines) plaquettes which are either fully filled, or one site is empty, see Eq.~\eqref{nPl}. }
\label{Patterns}
\end{figure}

The above probabilistic rules of the model are presented in Table~\ref{RulesTab}
and visualized in Fig.~\ref{Patterns}a. The kinetic process \eqref{ModDef}
can be considered as directed percolation on the (1+1) (space-time) lattice shown in Fig.~\ref{Patterns}; its absorbing (active) phase is
non-percolative (percolative) phase on the 2D lattice, respectively.
This model was introduced and
studied in \cite{Chitov:2015}. It can be considered as a modification  of the well-known model of the directed bond percolation (BDP)  \cite{Hinrichsen:2000}.

%
%
%%%%%%%%%%%%%%%%%%%%%%%%%%%%%%%%%%%%%%%%%%%%%%%%%%%%%%%%%%%%%%%%%%%%%%%%%%%%%%%%%%%%%%%%%%%%%%%%%%%%%%%%%%%%%%%%%%%%%%%%%%%%%%%%%%%%%%%%%%%%%%%%%%%%%%%%%%%%%%
%%%%%%%%%%%%%%%%%%%%%%%%%%%%%%%%%%%%%%%%%%%%%%%%%%%%%%%%%%%%%%%%%%%%%%%%%%%%
%
%

%
%%%%%%%%%%%%%%%%%%%%%%%%%%%%%%%%%%%%%%%%%%%%%%%%%%%%%%%%%%%%%%%%%%%%%%%%%%%%%%%%%%%%%%%%%%%%%%%%%%%%%%%%%%%%%%%%%%%%%%%%%%%%%%%%%%%%%%%%%%%%%%%%%%%%%%%%%%%%%%
\begin{table}[h]
\caption{Spreading rules of the model.}\label{RulesTab}
\begin{tabular}{|l|llll|ll|l|l|}
\hline
\begin{tabular}[c]{@{}l@{}}$n_{i-1,t}$\\ $n_{i,t}$\\ $n_{i+1,t}$\end{tabular}                   & \multicolumn{1}{l|}{\begin{tabular}[c]{@{}l@{}}1\\ 1\\ 1\end{tabular}} & \multicolumn{1}{l|}{\begin{tabular}[c]{@{}l@{}}1\\ 1\\ 0\end{tabular}} & \multicolumn{1}{l|}{\begin{tabular}[c]{@{}l@{}}0\\ 1\\ 1\end{tabular}} & \begin{tabular}[c]{@{}l@{}}0\\ 1\\ 0\end{tabular} & \multicolumn{1}{l|}{\begin{tabular}[c]{@{}l@{}}0\\ 0\\ 1\end{tabular}} & \begin{tabular}[c]{@{}l@{}}1\\ 0\\ 0\end{tabular} & \multicolumn{1}{c|}{\begin{tabular}[c]{@{}c@{}}1\\ 0\\ 1\end{tabular}} & \begin{tabular}[c]{@{}l@{}}0\\ 0\\ 0\end{tabular} \\ \hline
\begin{tabular}[c]{@{}l@{}}Probability\\ $n_{i,t+1} = 1$\end{tabular} & $p$                                                                      &                                                                        &                                                                        &                                                   & $q$                                                                      &                                                   & $q(2-q)$                                                  & $0$                                                  \\ \hline
\end{tabular}
\end{table}
%%%%%%%%%%%%%%%%%%%%%%%%%%%%%%%%%%%%%%%%%%%%%%%%%%%%%%%%%%%%%%%%%%%%%%%%%%%%%%%%%%%%%%%%%%%%%%%%%%%%%%%%%%%%%%%%%%%%%%%%%%%%%%%%%%%%%%%%%%%%%%%%%%%%%%%%%%%%%%
%

The absorbing-active phase transition is probed by the order parameter
\begin{equation}
\label{rho}
  \rho(t) = \frac{1}{N} \left\langle  \sum_{i=1}^{N} n_{i,t}  \right\rangle~,
%  \lim_{N \to \infty}
\end{equation}
where \(\langle \ldots \rangle\) denotes averaging over random trials, such that $\rho(\infty) =0$ in the absorbing phase, while in the active
phase $0<\rho(\infty) \leq 1$. In spite of the seemingly local form of $\rho(t)$, it detects nonlocal properties of the active
phase, giving simultaneously the probability of existence of a path (string) \cite{Chitov:2016} connecting active sites in the whole time interval $[0,t]$. This is a consequence of the fact that every filled site must have at least one ancestor at the preceding time step, as follows from the transfer probabilities of the model \eqref{ModDef}.

As known, coarse graining (e.g., the Kadanoff-Wilson scheme) implemented for a local order parameter cannot yield new critical points,
i.e., the coarse-grained and the original order parameters appear simultaneously. This is one of the foundations of the Wilson renormalization-group theory
of critical phenomena \cite{Wilson:1974}.
The same is not true for percolation: coarse graining implemented via construction of different percolative (geometric) patterns, as defined above,
engenders cascades of transitions, when distinct order parameters emerge within the active phase. It has been numerically confirmed for several
models \cite{Chitov:2015,Chitov:2016}.

We will study the critical properties of the model \eqref{ModDef} using the finite-size scaling analysis of the raw MC data.
For the DP class transition one needs three independent exponents \cite{Hinrichsen:2000}. In the active phase the order parameter near the absorbing-active transition scales as:
\begin{equation}
\label{beta}
  \rho(\infty) \propto (p - p_c)^{\beta}~,
\end{equation}
and similarly with  $p \leftrightarrow q$, depending on the chosen direction in the parametric plane $(p,q)$. The dynamical transition is characterized by the spatial  ($\xi_{\perp}$) and the temporal ($\xi_{\parallel}$) correlation lengths, diverging at the critical point as:
\begin{equation}
\label{nu}
   \xi_{\perp} \propto |p - p_c|^{-\nu_{\perp}}, \quad \xi_{\parallel}
   \propto |p - p_c|^{-\nu_{\parallel}} ~.
\end{equation}
Two correlation lengths $\xi_{\parallel} \sim \xi_{\perp}^z$ are related  through the dynamical exponent
\begin{equation}
\label{zet}
  z = \frac{\nu_{\parallel} }{ \nu_{\perp}}~.
\end{equation}
For a finite-size system the order parameter scales as \cite{Hinrichsen:2000}:
\begin{equation}
\label{rhot}
  \rho(t) \sim t^{-\alpha} f\left(\Delta t^{1/\nu_{\parallel}} , \frac{t^{1/z}}{N} \right)~,
\end{equation}
where $\Delta$ is the distance from the critical point, and the exponent
\begin{equation}
\label{ban}
   \alpha = \frac{\beta}{\nu_{\parallel}}~.
\end{equation}
determines the critical decay of $\rho(t)$ at $\Delta=0$ in the limit $N \to \infty$.

From analysis of \eqref{rhot} with various data sets, we infer the critical points and the indices $\alpha$, $\nu_\parallel$, and $z$, as explained in  the next section.  Once \(\alpha\) and $\nu_\parallel$ are found, the value of \(\beta\) is calculated from \eqref{ban}.

As an independent cross-check for the location of the critical points found from \eqref{rhot}, we also calculate the probability $P_{N,T}(p,q)$ that a connected cluster spanning from $t=0$ to $t=T$ occurs in the system of size $N \times T$.   $P_{N,T}(p,q)$ is simply given by the fraction of connected sites.
For a given value of $p$, all plots of $P_{N,T}(q)$ for different sizes $N \times T$ intersect at the critical point $q=q_c$,
thus $P_{N,T}(q)$ plays the role similar to the Binder cumulant for scaling analysis of percolation transition \cite{Binder:2019}.

%xxxxxxxxxxxxxxxxxxxxxxxxxxxxxxxxxxxxxxxxxxxxxxxxxxxxxxxxxxxxxxxxxxxxxxxxxxxxxx
%
%xxxxxxxxxxxxxxxxxxxxxxxxxxxxxxxxxxxxxxxxxxxxxxxxxxxxxxxxxxxxxxxxxxxxxxxxxxxxxx
\section{Results}\label{Res}
%%%%%%%%%%%%%%%%%%%%%%%%%%%%%%%%%%%%%%%%%%%%%%%%%%%%%%%%%%%%%%%%%%%%%%%%%%%%
%xxxxxxxxxxxxxxxxxxxxxxxxxxxxxxxxxxxxxxxxxxxxxxxxxxxxxxxxxxxxxxxxxxxxxxxxxxxxxx
%xxxxxxxxxxxxxxxxxxxxxxxxxxxxxxxxxxxxxxxxxxxxxxxxxxxxxxxxxxxxxxxxxxxxxxxxxxxxxx
%
%
%xxxxxxxxxxxxxxxxxxxxxxxxxxxxxxxxxxxxxxxxxxxxxxxxxxxxxxxxxxxxxxxxxxxxxxxxxxxxxx
%xxxxxxxxxxxxxxxxxxxxxxxxxxxxxxxxxxxxxxxxxxxxxxxxxxxxxxxxxxxxxxxxxxxxxxxxxxxxxx
%
\subsection{Active and dipole phases}\label{ADphases}
\begin{figure}[h]
%\vspace{-1em}
\centering{\includegraphics[width=8cm]{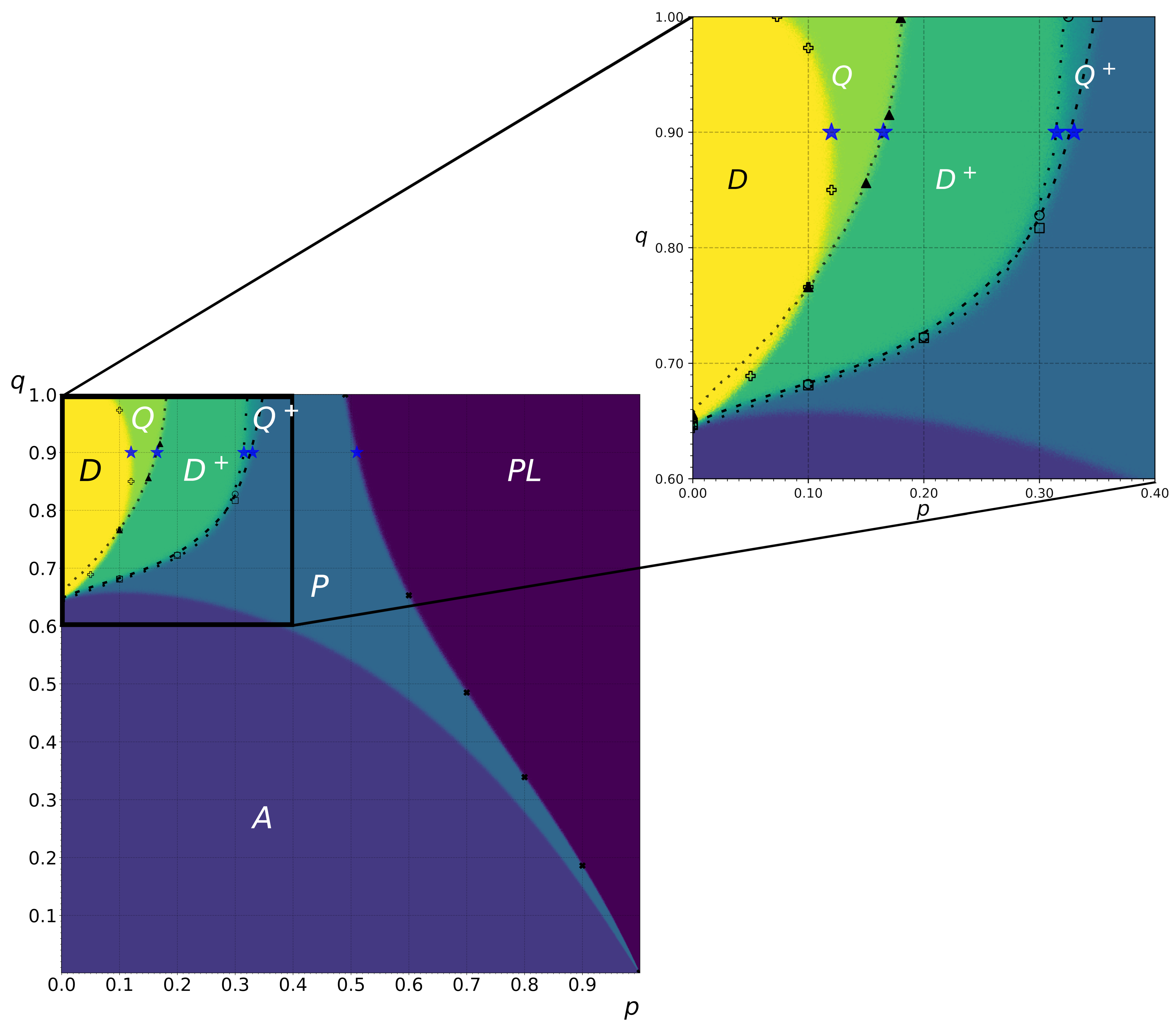}}%9.5
\caption{Phase diagram of the model. $P$ -  percolating phase without hidden patterns, $D$ ($D^+$) and  $Q$ ($Q^+$) - phases with dipole and quadrupole percolating patterns with nn (nn+nnn) bonds, respectively. $PL$ - plaquette phase. $A$ - absorbing state. Blue filled stars - points along the line $q=0.9$, where different percolating patterns and tower of order parameters are calculated. The upper left corner of the phase diagram (inside the black square) is zoomed for better view.}
\label{PDiag}
%\vspace{-1em}
\end{figure}

%xxxxxxxxxxxxxxxxxxxxxxxxxxxxxxxxxxxxxxxxxxxxxxxxxxxxxxxxxxxxxxxxxxxxxxxxxxxxxx
%xxxxxxxxxxxxxxxxxxxxxxxxxxxxxxxxxxxxxxxxxxxxxxxxxxxxxxxxxxxxxxxxxxxxxxxxxxxxxx
%
The absorbing-active transition and the transition within the active (percolating) phase into the phase with dipole percolation pattern were reported in the earlier work \cite{Chitov:2015}. The current results confirm those phase boundaries on the phase diagram in Fig.~\ref{PDiag} and the DP universality class of both transitions. The attributes of the active phase are standard \cite{Hinrichsen:2000}, below we recapitulate the definition of the dipole phase to make the presentation more self-contained and to better explain some subtle points of \cite{Chitov:2015}.

%
%%%%%%%%%%%%%%%%%%%%%%%%%%%%%%%%%%%%%%%%%%%%%%%%%%%%%%%%%%%%%%%%%%%%%%%%%%%%
%%%%%%%%%%%%%%%%%%%%%%%%%%%%%%%%%%%%%%%%%%%%%%%%%%%%%%%%%%%%%%%%%%%%%%%%%%%%
%%%%%%%%%%%%%%%%%%%%%%%%%%%%%%%%%%%%%%%%%%%%%%%%%%%%%%%%%%%%%%%%%%%%%%%%%%%%
%%%%%%%%%%%%%%%%%%%%%%%%%%%%%%%%%%%%%%%%%%%%%%%%%%%%%%%%%%%%%%%%%%%%%%%%%%%%
%

\begin{figure}[h]
\centering{\includegraphics[width=2cm]{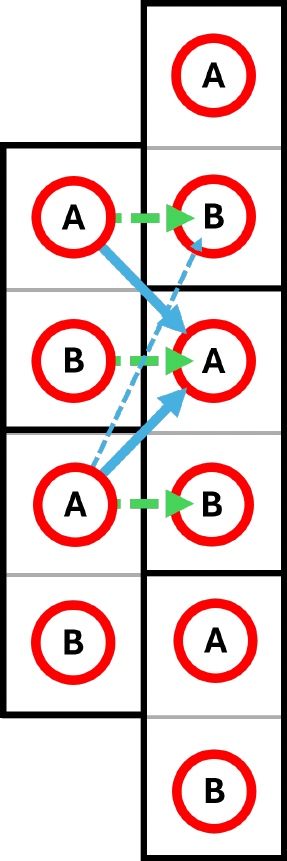}}
\caption{Two sublattices A/B of the original space-time lattice. Time direction is horizontal. Selection of dipoles shown in bold. The $q$ transfer probabilities to replicate a filled site A from its A-ancestors shown in solid light blue. Some of the $p$ transfer probabilities shown in dashed green. The interplay of $q-$ and $p-$ contributions allows to admit percolating bond between nnn dipoles, shown in dashed blue. }
\label{Fig0}
\end{figure}

To understand the simulation results given below, let us first get some intuitive understanding of the model. In the limit $p=0$ it is equivalent to the directed
bond percolation (BDP) with the percolating threshold $q=q_{\mathrm{BDP}} \approx 0.6447$. The BDP is taking place on two sublattices ($A$ and $B$) shown in Fig.~\ref{Fig0}. The two sublattices are correlated due to the term  $P(1|1,0,1)$. Moreover, since $P(1|1,0,1)>\{P(1|0,0,1)=P(1|1,0,0)\}$, the replication with
probability $q$ promotes the active phase with alternated filling $(..101010..)$ in the spatial and temporal directions. At the other end of the active phase, when $p=1$ ($q >0$), the stable state is a fully occupied lattice $(..1111..)$ in both directions. The above arguments suggest that at least one transition
could occur in the range  $p \in [0,1]$ to separate two active phases with distinct percolation patterns.

The above heuristic arguments are further supported by the mean-field solution of the master equation for the model \cite{Chitov:2015}.  The approximation based on the decoupling of the three-particle distribution function, predicts a continuous transition in the region of small $p$ inside the active phase, quite close to the
numerical $D$-line on the exact phase diagram shown in Fig.~\ref{PDiag}.
The transition separates a phase with uniform filling in the stationary state at $p>p_c(q)$ from the ``antiferromagnetic-like'' (AFM) phase with different subblattice occupations $(..n_A n_B n_A n_B..)$, $n_A \neq n_B$ in both directions at $p<p_c(q)$. In the limit $p \to 0$,  one of the sublattices stays empty $n_A \to 0$  (or $n_B \to 0$). The mean-field approximation leads to the spontaneous breaking of sublattice symmetry $A \leftrightarrow B$.
%%%%%%%%%%%%%%%%%%%%%%%%%%%%%%%%%%%%%%%%%%%%%%%%%%%%%%%%%%%%%%%%%%%%%%%%%%%%%%%%%%%%%%%%%%%%%%%%%%%%%%%%%%%%%%%%%%%%%%%%%%%%%%%%%%%%%%%%%%%%%%%%%%%%%%%%%%%%%%
%%%%%%%%%%%%%%%%%%%%%%%%%%%%%%%%%%%%%%%%%%%%%%%%%%%%%%%%%%%%%%%%%%%%%%%%%%%%%%%%%%%%%%%%%%%%%%%%%%%%%%%%%%%%%%%%%%%%%%%%%%%%%%%%%%%%%%%%%%%%%%%%%%%%%%%%%%%%%%
%

%
%%%%%%%%%%%%%%%%%%%%%%%%%%%%%%%%%%%%%%%%%%%%%%%%%%%%%%%%%%%%%%%%%%%%%%%%%%%
%%%%%%%%%%%%%%%%%%%%%%%%%%%%%%%%%%%%%%%%%%%%%%%%%%%%%%%%%%%%%%%%%%%%%%%%%%%
\begin{figure*}[h]
\begin{subfigure}{0.4\textwidth}
    \includegraphics[width=7.2cm]{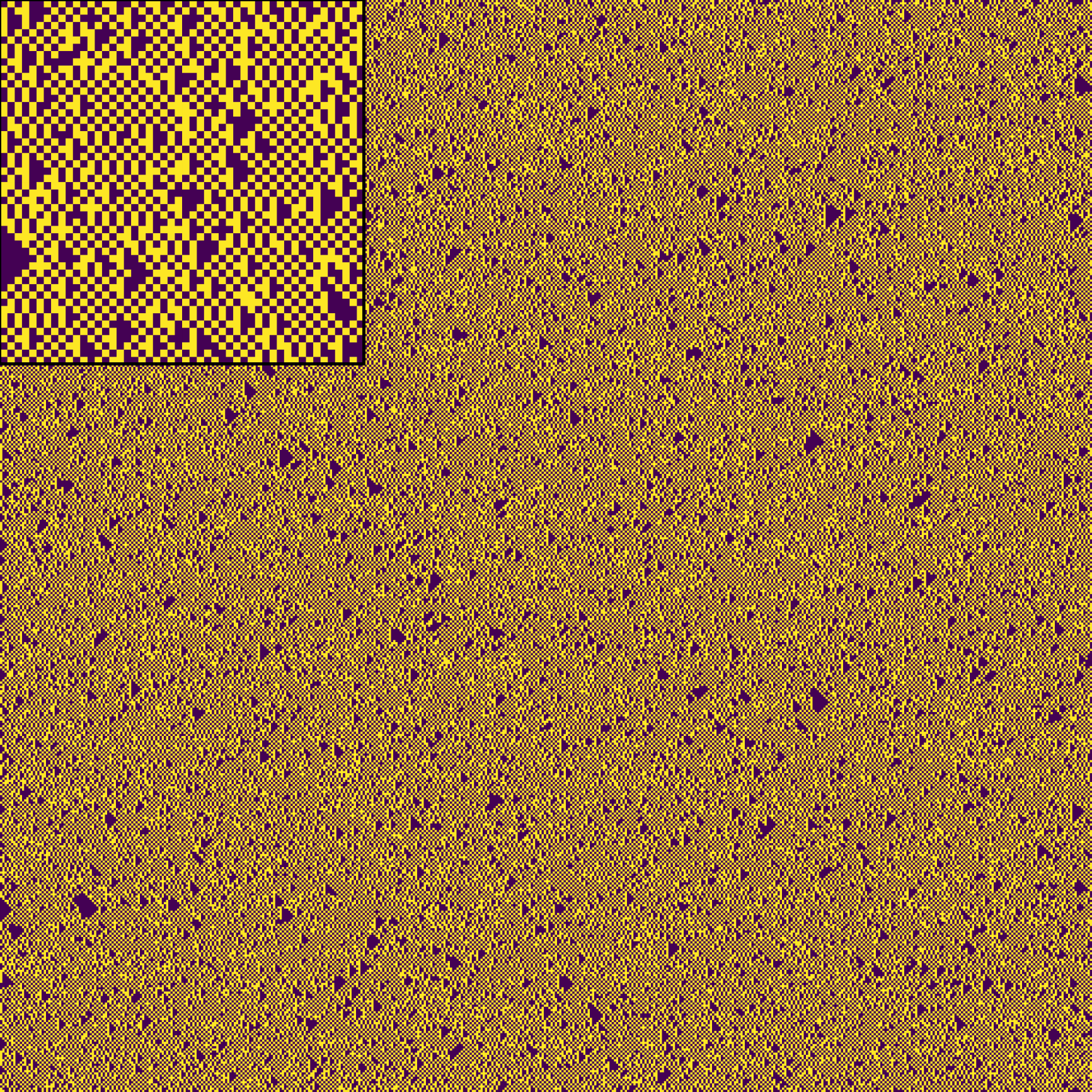}
    \caption{}
    \label{fig:1a}
  \end{subfigure}%
  \hspace{0.03\textwidth}
\begin{subfigure}{0.4\textwidth}
    \includegraphics[width=7.2cm]{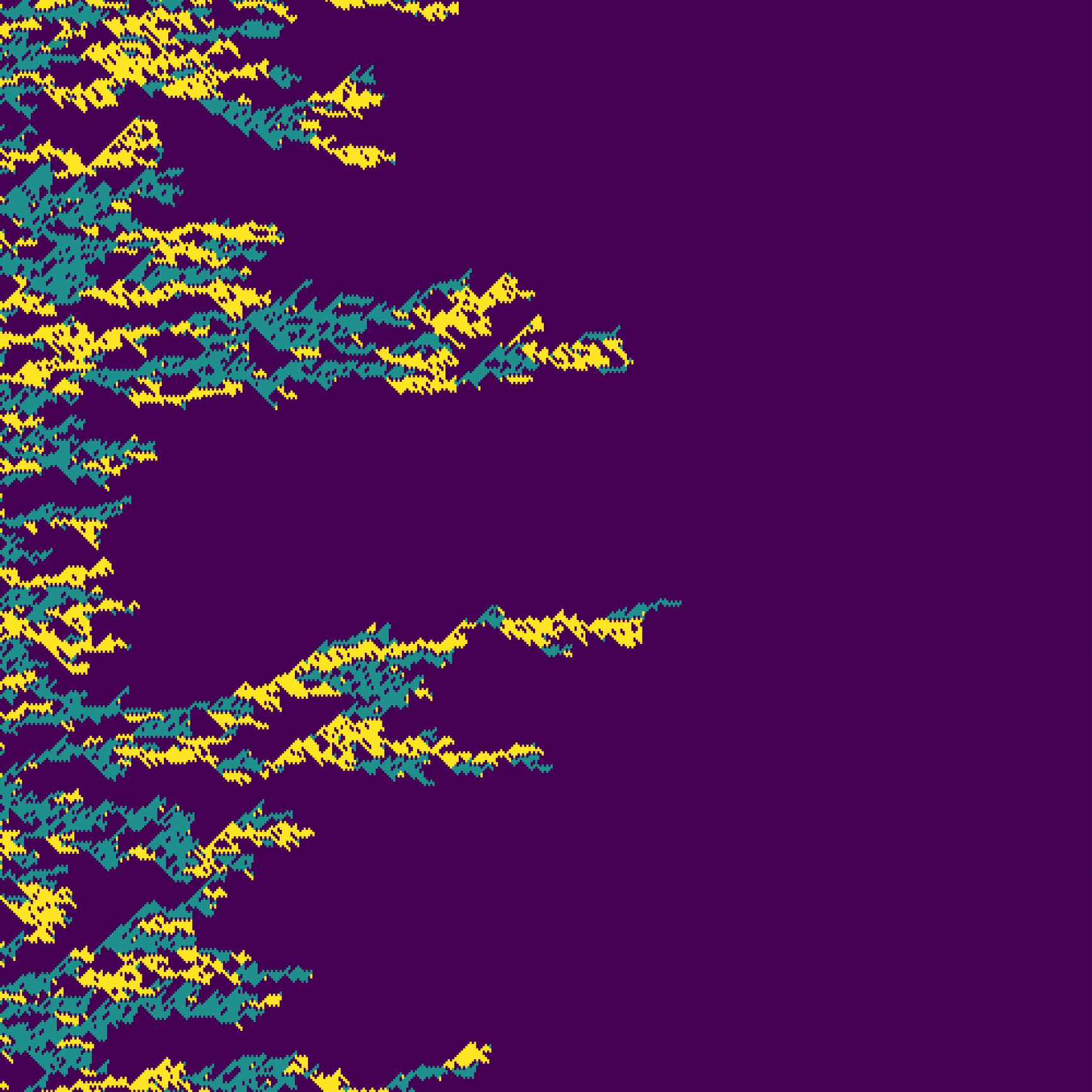}
    \caption{}
    \label{fig:1b}
  \end{subfigure}%

\begin{subfigure}{0.4\textwidth}
\includegraphics[width=7.2cm] {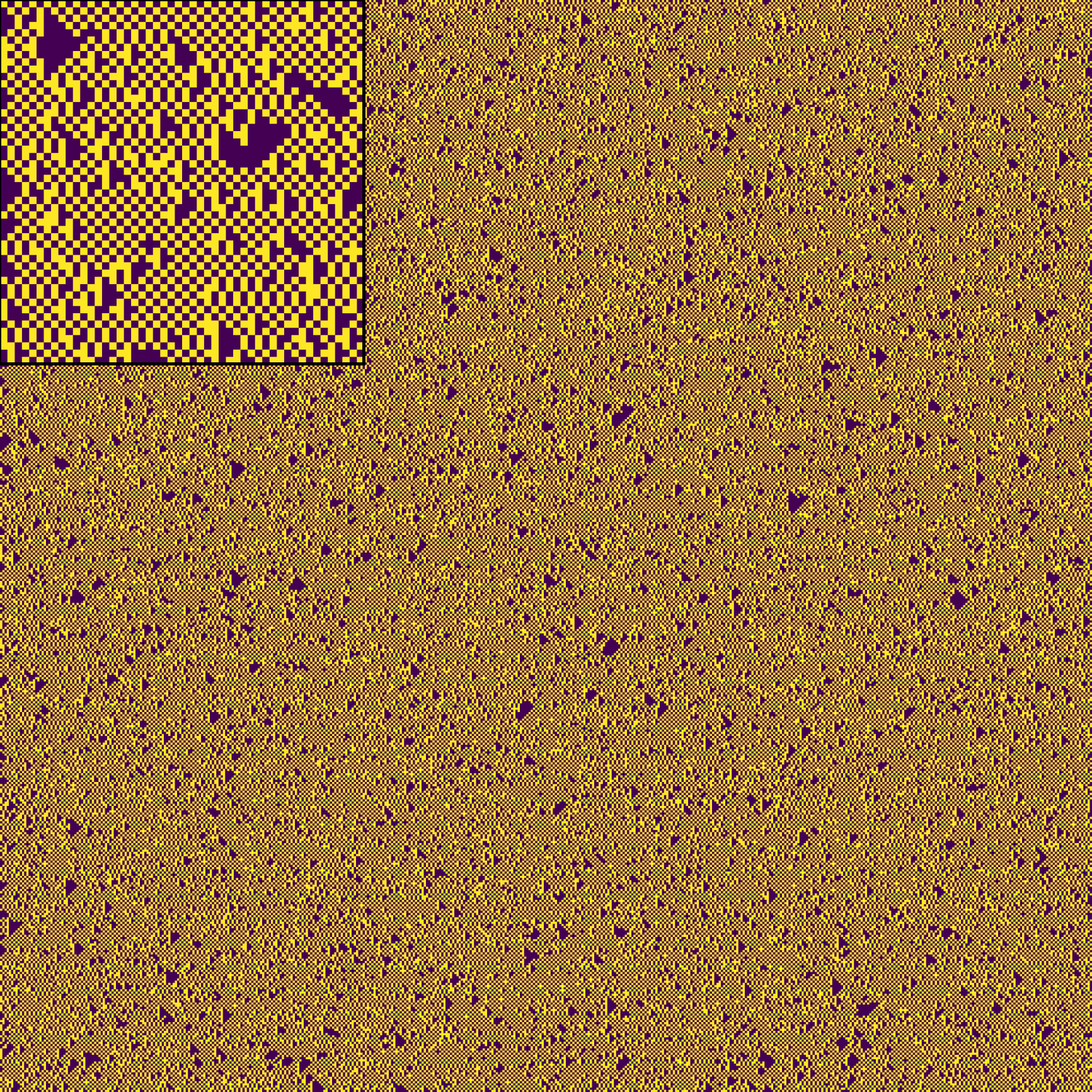}
    \caption{} \label{fig:1c}
  \end{subfigure}%
  \hspace{0.03\textwidth}
\begin{subfigure}{0.4\textwidth}
    \includegraphics[width=7.2cm]{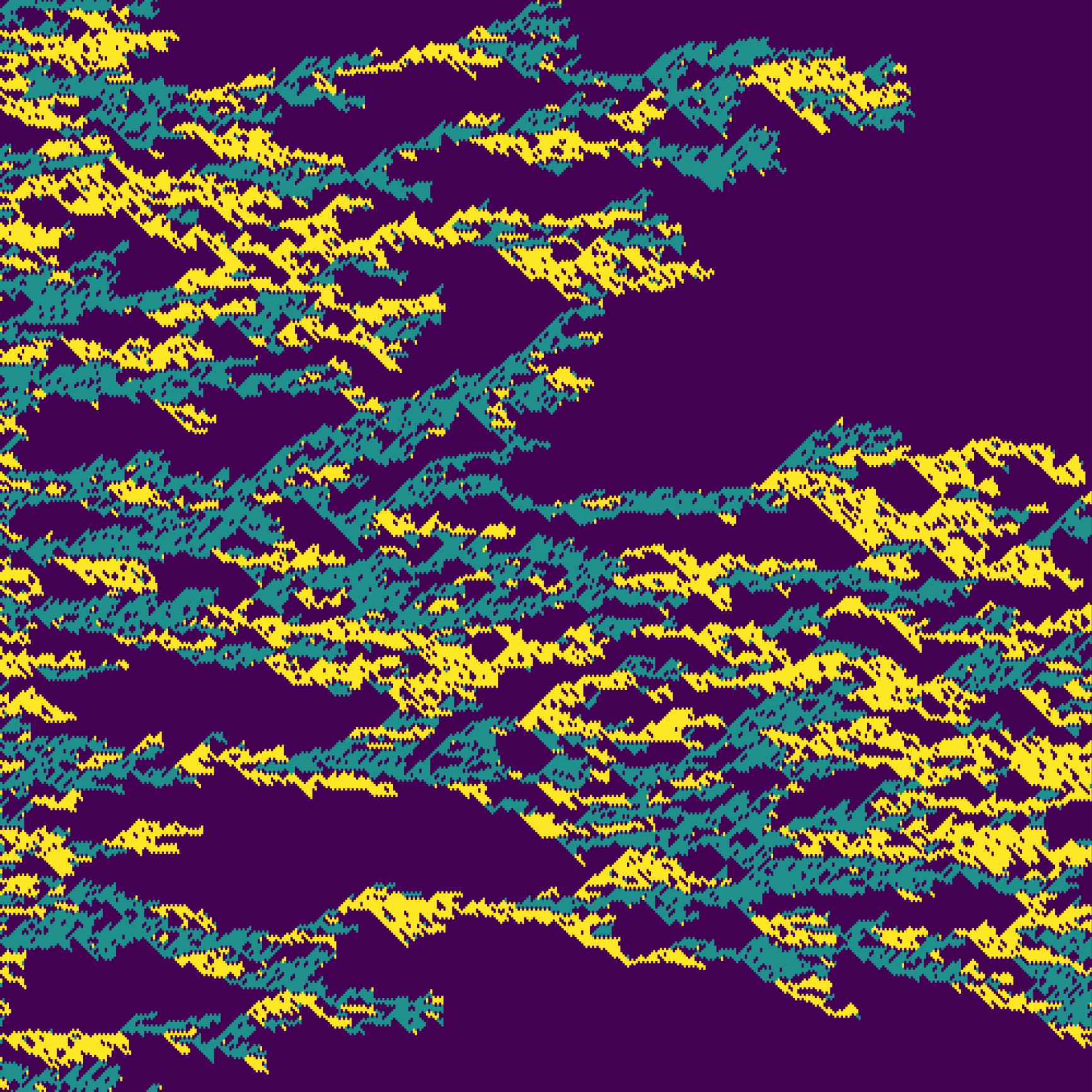}
    \caption{}
    \label{fig:1d}
  \end{subfigure}%
\caption{$D$-patterns: (a,c) MC data ($N=500$, $T=500$) for two configurations of the percolating phase at $q=0.9$; (a) $p=0.14$, (c) $p=0.12$; critical point of the transition into $D$ phase $p_c(q) \approx 0.121$; ochre/purple dots correspond to the filled/empty sites of the original lattice. Zoomed fragments of raw data shown in insets. (b,d) The connected dipole (nn) patterns constructed from the data shown in (a,c); (b) $p>p_c(q)$, dipole pattern is not percolative; (d) $p<p_c(q)$, the system is in the percolating $D$ phase; yellow/teal dots correspond to connected $\pm 1$ dipoles, respectively, residing on the sites of the dipole lattice; the dark purple background corresponds to the sites without dipoles or to disconnected dipoles without nn ancestors. }
\label{fig:dipole_pattern_example}
\end{figure*}
%%%%%%%%%%%%%%%%%%%%%%%%%%%%%%%%%%%%%%%%%%%%%%%%%%%%%%%%%%%%%%%%%%%%%%%%%%%
%%%%%%%%%%%%%%%%%%%%%%%%%%%%%%%%%%%%%%%%%%%%%%%%%%%%%%%%%%%%%%%%%%%%%%%%%%%
%

The MC simulations disprove the mean-field prediction of the local AFM-like order with the occupancy disproportionation, but the mean field turns out to be very instrumental to reveal a true and more subtle transition between the percolative patterns in the active phase. A simply percolating phase (we will call it just ``percolated'' ($P$) in the following) is detected by the non-vanishing standard order parameter $\rho (\infty)$ \eqref{rho}.
To reveal the other percolating phase with its own order parameter, we will coarse grain the sites of the original lattice as follows: for a given 2D ($N \times T$) MC data set we analyse the connectivity of the dipoles residing on the nodes of the dipole lattice shown in  Fig.~\ref{Patterns}b.
The occupation numbers on the sites of the coarse-grained (dipole) lattice for the four possible dipole configurations are defined as follows:
\begin{eqnarray}
  \tilde n_{i,t} &=& \left(
          \begin{array}{c}
            1 \\
            0 \\
          \end{array}
        \right)=1,~~
  \tilde n_{i,t} = \left(
          \begin{array}{c}
            0 \\
            1 \\
          \end{array}
        \right)=-1,~~
  \nonumber \\
   \tilde n_{i,t} &=& \left(
          \begin{array}{c}
            0 \\
            0 \\
          \end{array}
        \right)=
  \left(
          \begin{array}{c}
            1 \\
            1 \\
          \end{array}
        \right)=0.~~
\label{Dns}
\end{eqnarray}
The percolation (connectivity) on the dipole lattice can be allowed between the nearest neighboring (nn) and the next-nearest neighboring (nnn) active sites. The site on the coarse grained lattice is active if a dipole resides on it, $\tilde n_{i,t} = \pm 1$.\footnote{We kept the distinction between $\tilde n_{i,t} = 1$ and $\tilde n_{i,t} = -1$  for $D,D^+$ phases mostly for cross-checking purposes, to make sure that there is no symmetry breaking between two types ($\pm$) of connected clusters and they appear simultaneously, as visualized in the figures shown below. For the scaling analysis, phase boundaries, etc, such distinction is unnecessary, and the absolute value $|\tilde n_{i,t}|$ was used.
The same applies for the $Q,Q^+$ phases, see definition \eqref{Qns} \label{PlusMinus}}
A more loosed connectivity (percolation) between nnn dipoles is introduced to account for the correlations  occurring due to $p$-replication probabilities, as shown in Fig.~\ref{Fig0}. In this work we study two dipole phases: $D$ phase is the one where the percolation only between nn dipoles is accounted for; and in the $D^+$ phase the percolation patterns include bonds between the nn and nnn active sites. The latter phase was first detected in \cite{Chitov:2015}.

To determine the lines of phase transitions and critical indices we run MC simulations
according to the rules \eqref{ModDef} for different sets of sizes $N$ and time steps $T$,
averaging each set over 1000 trials.
The line of the absorbing-percolating phase transition, see Fig.~\ref{PDiag},
reproduces exactly the result of  \cite{Chitov:2015} along with the critical indices of the DP universality class.

\begin{table}[h]
\centering
\caption{Critical points $q_{c}$ and critical indices for the transition into $D$ phase for several values of $p$. Scaling analysis presented in Fig.~\ref{fig:DSC} yields parameters shown in bold.}
{%
\begin{tabular}{|l|l|l|l|l|l|l|}
\hline
$p$ & $q_c$ & $\alpha$ & $\nu_{||}$ & z & $\beta = \alpha\nu_{||}$ & $\nu_{\perp} = \nu_{||} / z $ \\ \hline
0.0000 & 0.6444 & 0.1595 & 1.72 & 1.58 & 0.27(4) & 1.08(8) \\ \hline
0.0500 & 0.6896 & 0.1595 & 1.72 & 1.56 & 0.27(4) & 1.10(2) \\ \hline
\textbf{0.1000} & \textbf{0.7662} & \textbf{0.1595} & \textbf{1.72} & \textbf{1.56} & \textbf{0.27(4)} & \textbf{1.10(2)} \\ \hline
0.2002 & 0.8500 & 0.1595 & 1.72 & 1.56 & 0.27(4) & 1.10(2) \\ \hline
0.1000 & 0.9735 & 0.1595 & 1.72 & 1.56 & 0.27(4) & 1.10(2) \\ \hline
0.0736 & 1.0000 & 0.1590 & 1.72 & 1.57 & 0.27(3) & 1.09(5) \\ \hline
\end{tabular}%
}
\label{Tab:Dtr}
\end{table}
%%%%%%%%%%%%%%%%%%%%%%%%%%%%%%%%%%%%%%%%%%%%%%%%%%%%%%%%%%%%%%%%%%%%%%%%%%%%

To cross-check the codes and procedure implemented in this work, we proceeded first with the detection of the $D^+$ phase inside the active $P$ phase.  We calculate the dipole occupation $\tilde n$ \eqref{Dns} on the sites of the dipole lattice for each set of 2D MC data ($n$) obtained via  the process \eqref{ModDef} on the original lattice. The appearance of the dipole percolation
is signalled by non-vanishing concentration of the active connected dipoles $\tilde \rho(T) \neq 0$, where  $\tilde \rho(t)$ is defined analogously to \eqref{rho}. The selection of the active connected dipoles is made iteratively at each time step: The dipole occupation \eqref{Dns} is calculated from the MC results at $t=0$ and $t=1$. Among the active dipoles $\tilde n_{i,1} = \pm 1$ only those who have active ancestors, i.e.,  connected by the nn and/or nnn bonds to $\tilde n_{i,0} = \pm 1$, are retained. The disconnected active dipoles are discarded, $\tilde n_{i,1} \mapsto 0$. The procedure repeats until we reach $t=T$. The analysis of the $D^+$ phase yields the results in a complete agreement with those reported in earlier work \cite{Chitov:2015}, some data and results are given in Appendix \ref{AppB}.

To analyse the $D$ phase, which was not reported before,  we follow the same steps as above, with the only difference that the active connected dipoles at the time step $t$ are selected from those who have active ancestors at $t-1$, connected by the nn bonds \textit{only}.
In Fig.~\ref{fig:dipole_pattern_example} we show the raw MC results for the original occupancies $n$ on two sides from the critical line of the $D$ phase. The percolation patterns of the $P$ phase (a,c) are visually indistinguishable in both cases, while selection of the dipole percolation pattern with the nn connectivity (nn bonds) demonstrates clearly that the raw patterns are separated by a percolation transition into the $D$ phase.

\begin{figure*}[h]
\begin{subfigure}{0.4\textwidth}
    \includegraphics[width=7.2cm]{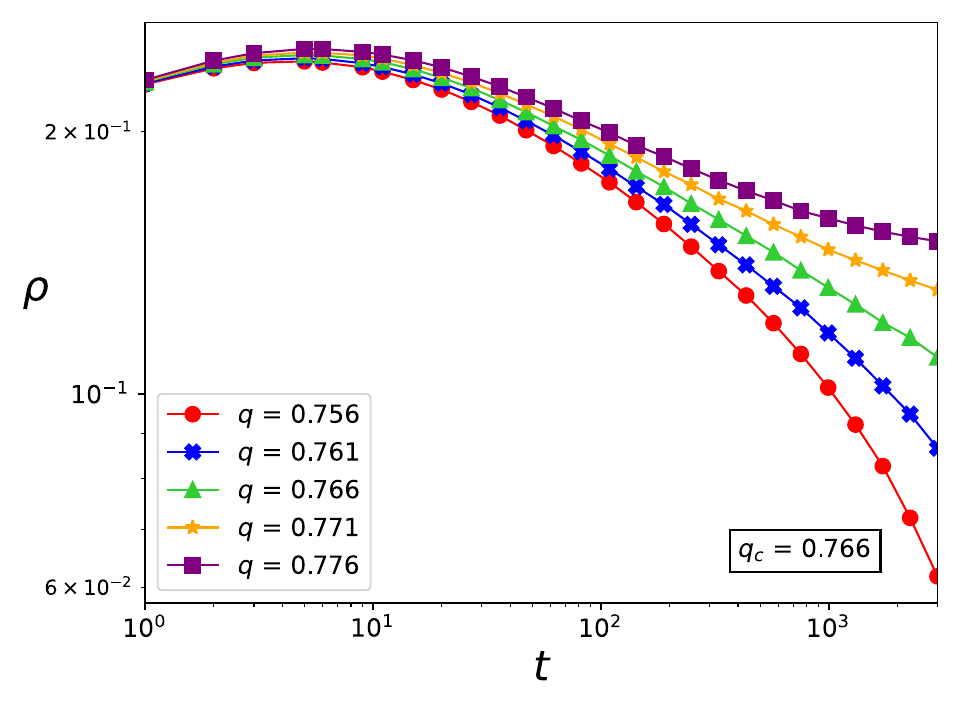}
    \caption{} \label{fig:1a}
  \end{subfigure}%
\begin{subfigure}{0.4\textwidth}
    \includegraphics[width=7.2cm]{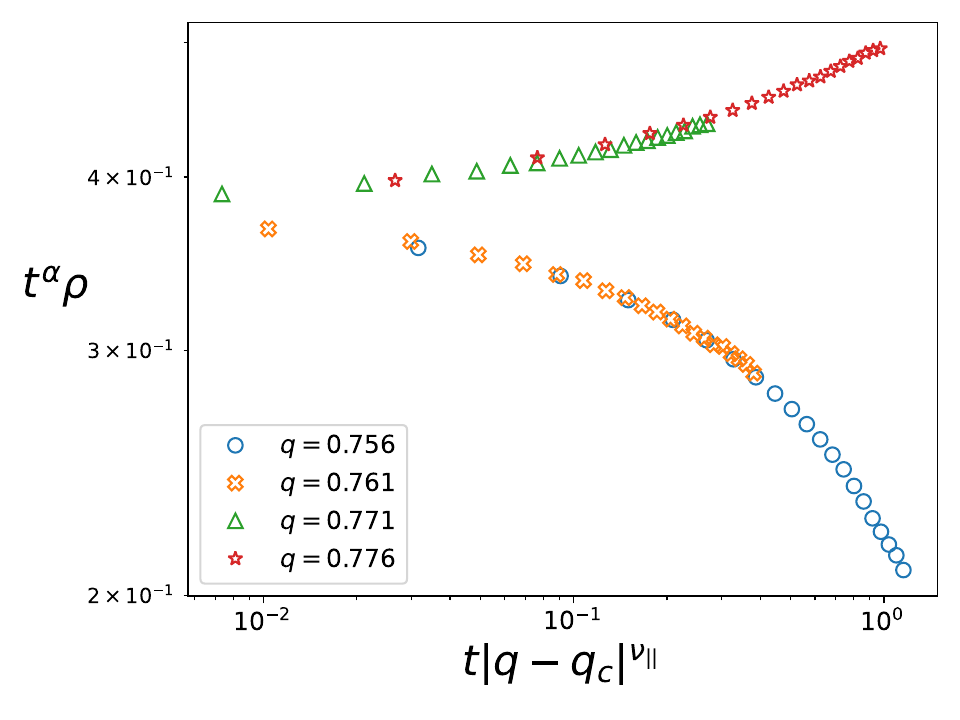}
    \caption{} \label{fig:1b}
  \end{subfigure}%

\begin{subfigure}{0.4\textwidth}
\hspace*{-6mm}
\includegraphics[width=7.4cm]
    {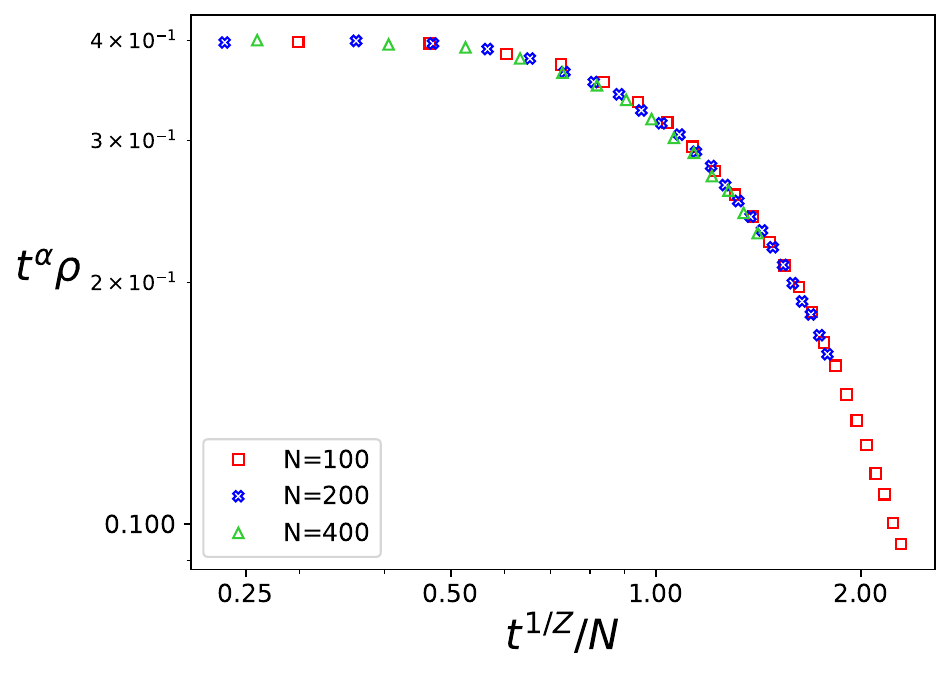}
    \caption{} \label{fig:1c}
  \end{subfigure}%
\begin{subfigure}{0.4\textwidth}
    \includegraphics[width=7.2cm]{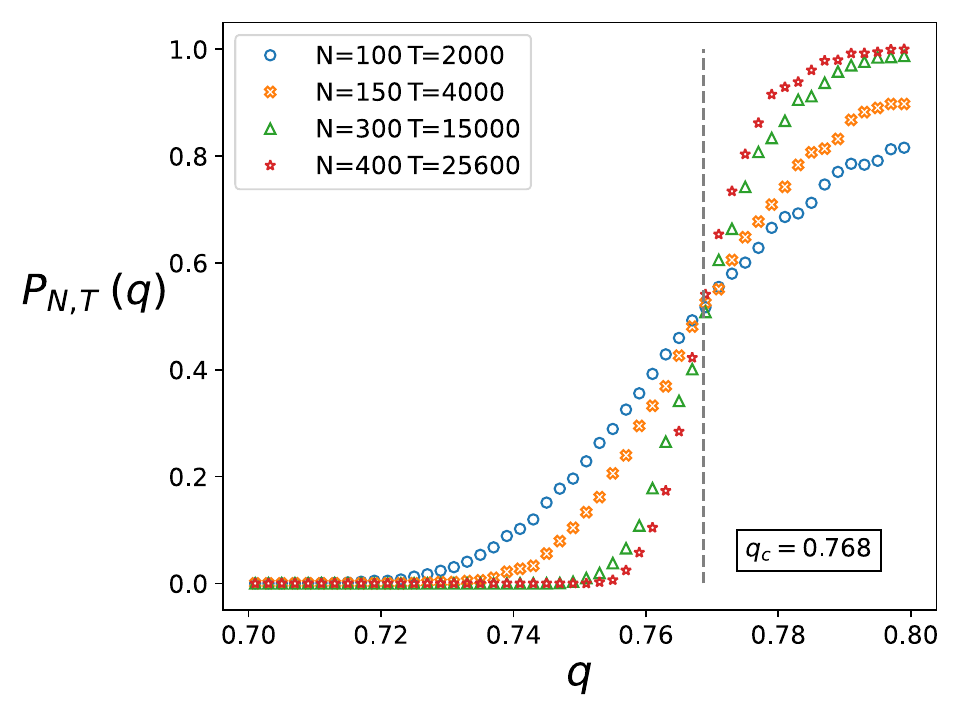}
    \caption{} \label{fig:1d}
  \end{subfigure}%
\caption{$D$ phase: (a) Density of the connected (nn) dipoles $\rho(t)$ for $p=0.1$ and series of $q$ near $q_c \approx 0.7662$. The system size is $N=3000$ and $T=100000$. The middle line corresponds to the critical slowing down $\rho(t) \propto t^{- \alpha}$ with $\alpha \approx 0.1595$.  (b) Collapse of the curves from (a) onto a single scaling function. Fitting gives the values of $q_c$ and $\nu_\parallel$. (c) Collapse of the order parameter relaxation $t^{\alpha}\rho(t)$ at the critical point  for different sizes ($N=100;~200;~400$) yields the critical index $z$.  (d) Fraction of connected dipoles $P_{N,T}(q)$ for different sizes $N \times T$, all curves intersect at  $q_c \approx  0.768$. The critical values of $q_c$ obtained from (a), (b) and (d) agree within $\sim 0.01 \% $  }
\label{fig:DSC}
%\vspace{-10em}
%\vspace{25cm}
\end{figure*}

In Fig.~\ref{fig:DSC}~(a) $\rho(t)$is plotted against $t$ on a double logarithmic scale.\footnote{From now on we keep tildes for the definitions of the coarse-grained occupation numbers, but drop them in graphs, etc, to unclutter notations.}
From collapse of the scaling functions for the system of different sizes shown in  Fig.~\ref{fig:DSC}~(b,c), the indices $\nu_{\parallel}$ and $z$ are obtained. The location of the critical point is also confirmed by the plots of the fraction of connected dipoles $P_{N,T}(q)$ for different sizes $N \times T$, shown in Fig.~\ref{fig:DSC}~(d).
They all intersect at $q=q_c$. The boundary of the $D$ phase is shown in Fig.~\ref{PDiag};
numerical results for several critical points $q_{c}$ and critical indices for the transition into this phase are collected in Table~\ref{Tab:Dtr}. The numbers confirm that this transition belongs to the DP universality class.

%
%
%
%%%%%%%%%%%%%%%%%%%%%%%%%%%%%%%%%%%%%%%%%%%%%%%%%%%%%%%%%%%%%%%%%%%%%%%%%%%
%%%%%%%%%%%%%%%%%%%%%%%%%%%%%%%%%%%%%%%%%%%%%%%
%
%
%

%
%
%xxxxxxxxxxxxxxxxxxxxxxxxxxxxxxxxxxxxxxxxxxxxxxxxxxxxxxxxxxxxxxxxxxxxxxxxxxxxxx
%xxxxxxxxxxxxxxxxxxxxxxxxxxxxxxxxxxxxxxxxxxxxxxxxxxxxxxxxxxxxxxxxxxxxxxxxxxxxxx
\subsection{Quadrupole phases}\label{Qphases}
%xxxxxxxxxxxxxxxxxxxxxxxxxxxxxxxxxxxxxxxxxxxxxxxxxxxxxxxxxxxxxxxxxxxxxxxxxxxxxx
%xxxxxxxxxxxxxxxxxxxxxxxxxxxxxxxxxxxxxxxxxxxxxxxxxxxxxxxxxxxxxxxxxxxxxxxxxxxxxx
%
%
The distinct percolation landscapes of the model in two limiting cases $p \to 0$ and $p \to 1$, discussed in the previous section, are suggestive to propose existence of other percolating phases with new patterns to be called quadrupole. We construct a new percolative backbone on the coarse grained lattice made out of $2\times2$ plaquettes (bold lines in  Fig.~\ref{Patterns}c) of the original lattice sites. The nodes of the coarse-grained lattice are placed at the centers of the plaquettes, see Fig.~\ref{Patterns}c.
%
%
%%%%%%%%%%%%%%%%%%%%%%%%%%%%%%%%%%%%%%%%%%%%%%%%%%%%%%%%%%%%%%%%%%%%%%%%%%%%
%%%%%%%%%%%%%%%%%%%%%%%%%%%%%%%%%%%%%%%%%%%%%%%%%%%%%%%%%%%%%%%%%%%%%%%%%%%%
\begin{figure*}[h]
\begin{subfigure}{0.4\textwidth}
    \includegraphics[width=7.2cm]{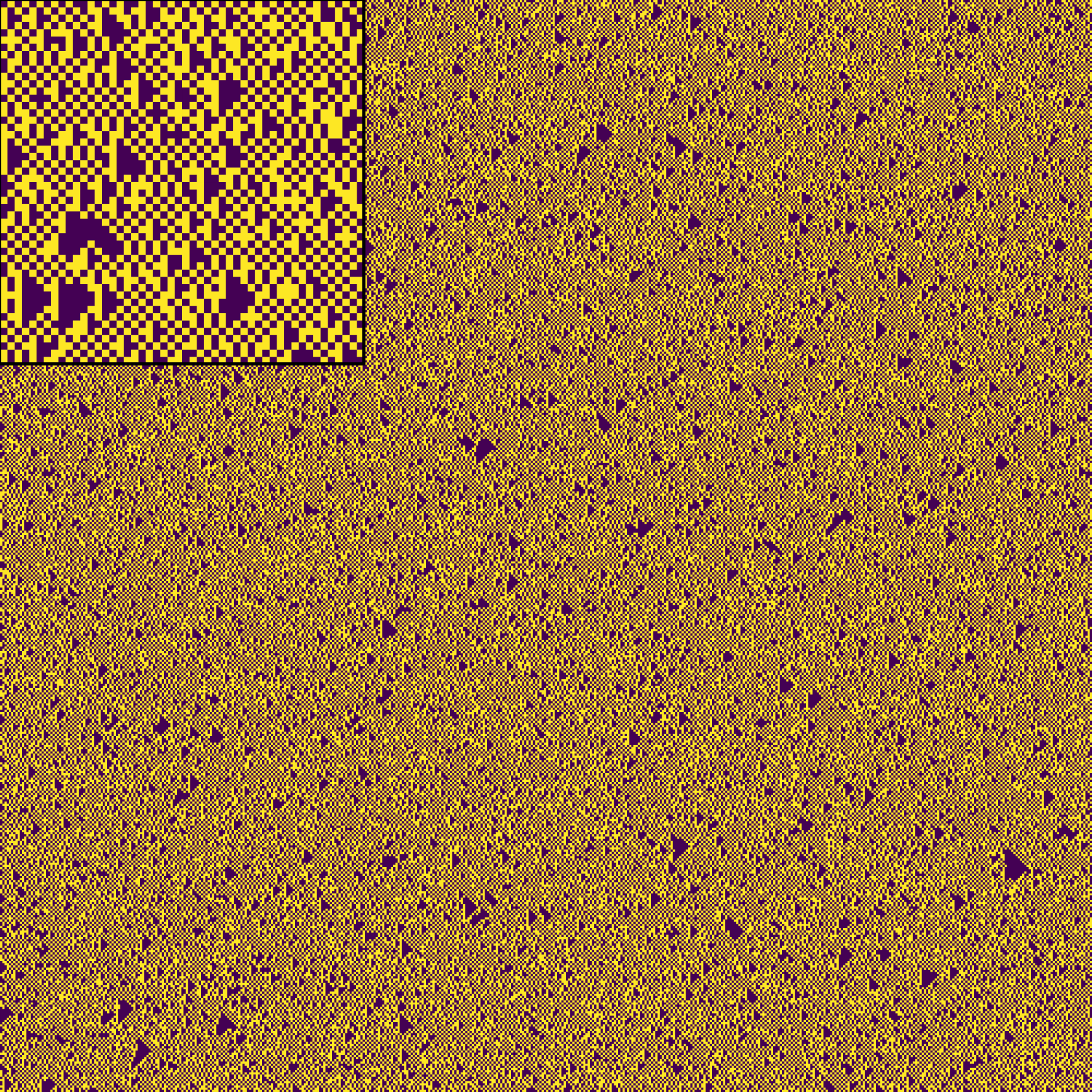}
    \caption{}
    \label{fig:1a}
  \end{subfigure}%
    \hspace{0.03\textwidth}
\begin{subfigure}{0.4\textwidth}
    \includegraphics[width=7.2cm]{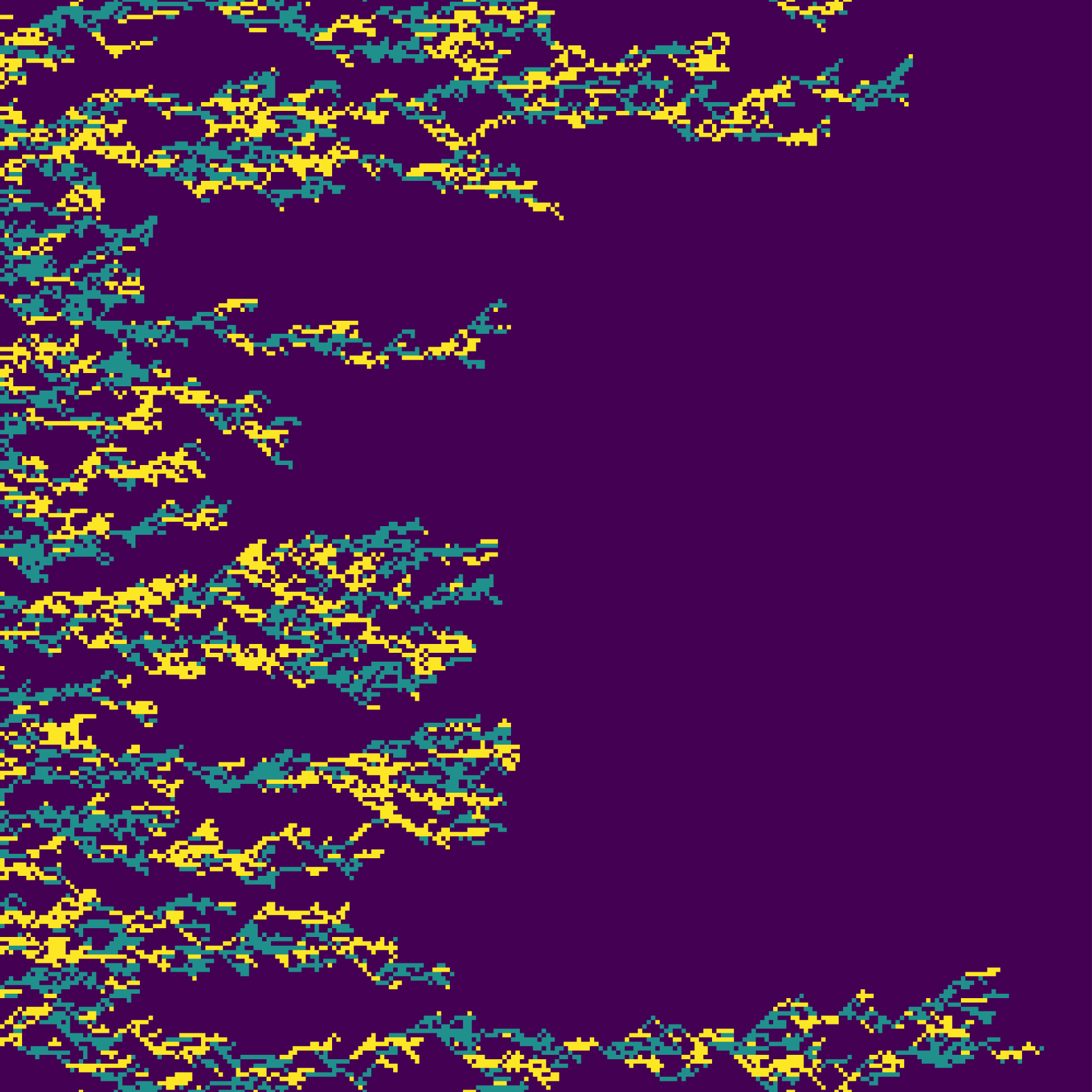}
    \caption{}
    \label{fig:1b}
  \end{subfigure}%

\begin{subfigure}{0.4\textwidth}
\includegraphics[width=7.2cm] {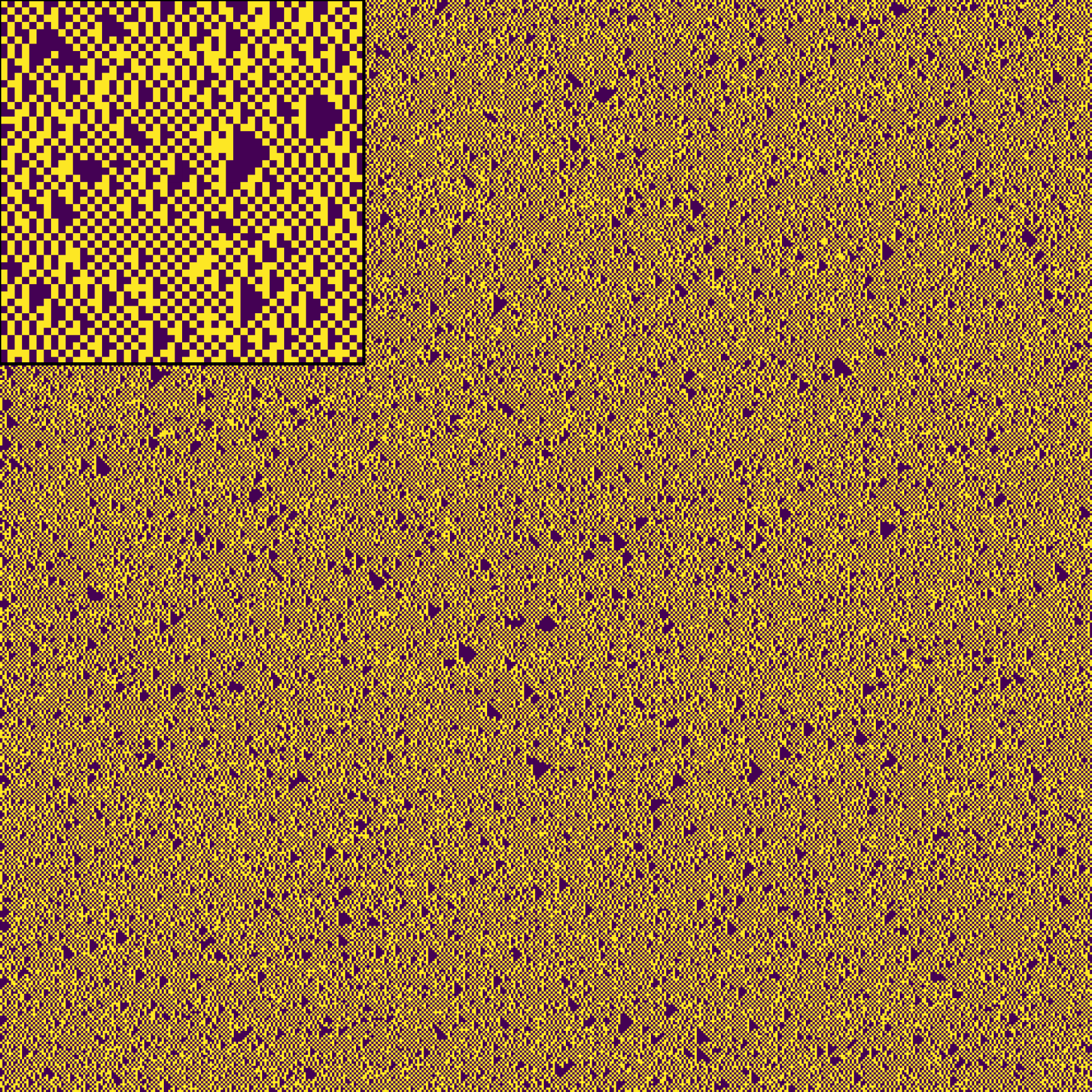}
    \caption{} \label{fig:1c}
  \end{subfigure}%
    \hspace{0.03\textwidth}
\begin{subfigure}{0.4\textwidth}
    \includegraphics[width=7.2cm]{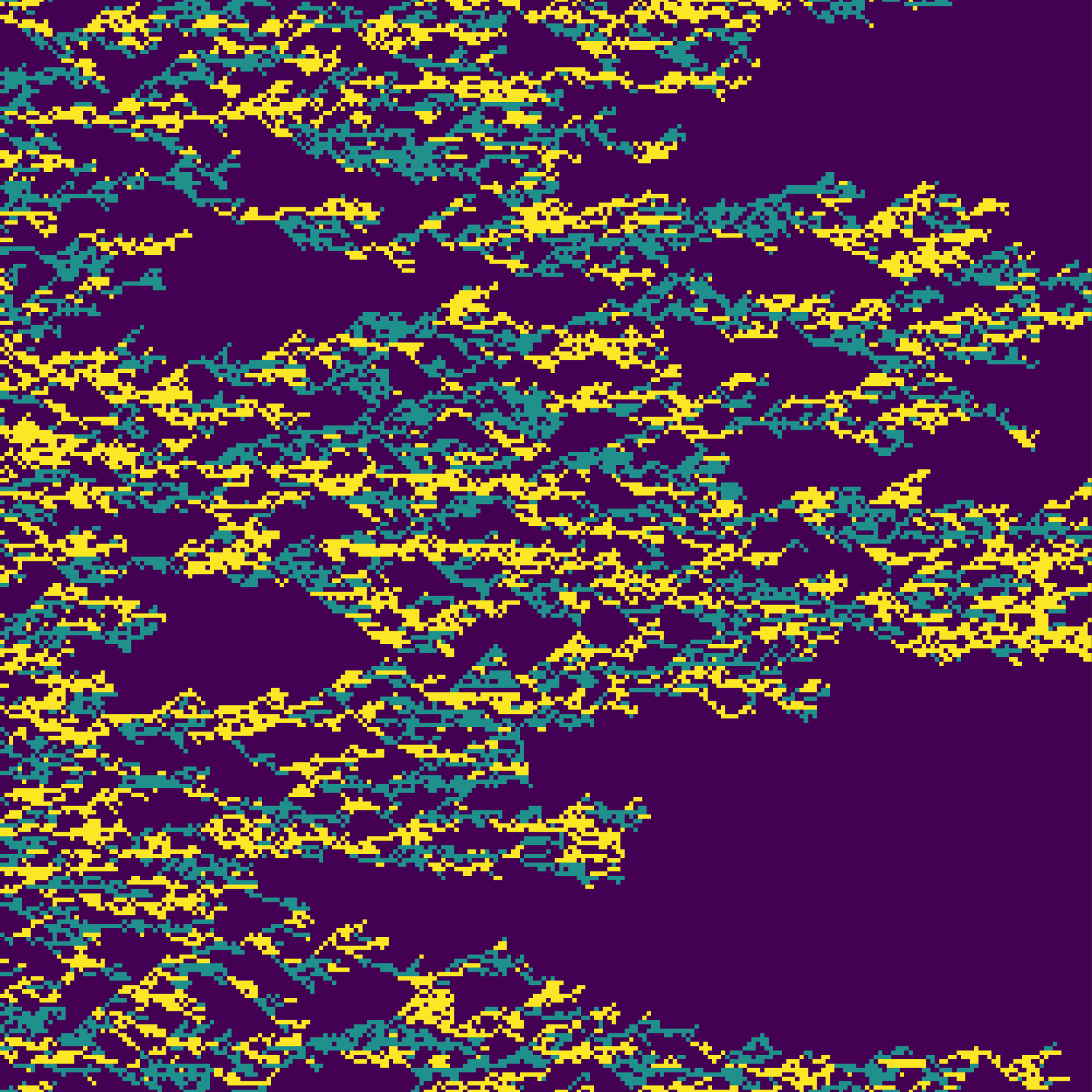}
    \caption{}
    \label{fig:1d}
  \end{subfigure}%
\caption{$Q$-patterns: (a,c) MC data ($N=500$, $T=500$) for two configurations of the percolating phase at $q=0.9$; (a) $p=0.18$, (c) $p=0.16$; critical point of the transition into $Q$ phase $p_c(q) \approx 0.167$; ochre/purple dots correspond to the filled/empty sites of the original lattice. Zoomed fragments of raw data shown in insets. (b,d) The connected quadrupole (nn) patterns constructed from the data shown in (a,c); (b) $p>p_c(q)$, quadrupole pattern is not percolative; (d) $p<p_c(q)$, the system is in the percolating $Q$ phase; yellow/teal dots correspond to connected $\pm 1$ quadrupoles, respectively, residing on the sites of the quadrupole lattice; the dark purple background corresponds to the sites without quadrupoles or to disconnected quadrupoles without nn ancestors.}
\label{fig:quadrupole_pattern_example}
\end{figure*}
%
%
%%%%%%%%%%%%%%%%%%%%%%%%%%%%%%%%%%%%%%%%%%%%%%%%%%%%%%%%%%%%%%%%%%%%%%%%%%%%
%%%%%%%%%%%%%%%%%%%%%%%%%%%%%%%%%%%%%%%%%%%%%%%%%%%%%%%%%%%%%%%%%%%%%%%%%%%%

The quadrupole occupation numbers on the nodes of the plaquette lattice are defined as:
\begin{eqnarray}
   \tilde n_{i,t} &=&
   \left(
     \begin{array}{cc}
       1 & 0 \\
       0 & 1 \\
     \end{array}
   \right)=1,~~
    \tilde n_{i,t} =
   \left(
     \begin{array}{cc}
       0 & 1 \\
       1 & 0 \\
     \end{array}
   \right)=-1,  \nonumber \\
  \tilde n_{i,t} &=& 0, ~~\mathrm{for~all~ other~ confugurations.}
\label{Qns}
\end{eqnarray}
The site is called active if $\tilde n_{i,t}= \pm 1$. (See the footnote\textsuperscript{\ref{PlusMinus}} for more details on this point.)
Then the procedure and analysis follow the same steps described in detail in the previous subsection. Again, we select two quadrupole percolative patterns: $Q^+$ where percolation is allowed between nn and nnn active sites and $Q$ where percolation is allowed between nn sites only.

The examples of the raw MC data on two sides from the transition are shown in Fig.~\ref{fig:quadrupole_pattern_example}, along the extracted $Q$ patterns. The latter clearly demonstrate occurrence of transition with the quadrupole percolative pattern. Some examples of the finite-size scaling analysis for the $Q$-phase transition are given in Fig.~\ref{QSC}, Appendix \ref{AppA}. The numerical results for several critical points and critical indices are presented in Table~\ref{Tab:Qtr},
Appendix \ref{AppA}.

Similar numerical results for the $Q^+$ phase are collected in Appendix \ref{AppB}.
The numbers confirm that both transitions belongs to the DP universality class.
The boundaries of the $Q$ and $Q^+$ phases are shown in Fig.~\ref{PDiag}.

%
%
%xxxxxxxxxxxxxxxxxxxxxxxxxxxxxxxxxxxxxxxxxxxxxxxxxxxxxxxxxxxxxxxxxxxxxxxxxxxxxx
%xxxxxxxxxxxxxxxxxxxxxxxxxxxxxxxxxxxxxxxxxxxxxxxxxxxxxxxxxxxxxxxxxxxxxxxxxxxxxx
%xxxxxxxxxxxxxxxxxxxxxxxxxxxxxxxxxxxxxxxxxxxxxxxxxxxxxxxxxxxxxxxxxxxxxxxxxxxxxx
%
\subsection{Plaquette phase}\label{PLphase}
%xxxxxxxxxxxxxxxxxxxxxxxxxxxxxxxxxxxxxxxxxxxxxxxxxxxxxxxxxxxxxxxxxxxxxxxxxxxxxx
%xxxxxxxxxxxxxxxxxxxxxxxxxxxxxxxxxxxxxxxxxxxxxxxxxxxxxxxxxxxxxxxxxxxxxxxxxxxxxx
%xxxxxxxxxxxxxxxxxxxxxxxxxxxxxxxxxxxxxxxxxxxxxxxxxxxxxxxxxxxxxxxxxxxxxxxxxxxxxx
%
%
As explained above, in the limit $p \to 1$ the stable state of the model is a fully occupied lattice in both (space-time) directions. From the above results  we expect an intermediate transition between $D,Q,PL$
patterns and a fully occupied state. For this end we consider again the coarse grained plaquette lattice shown in  Fig.~\ref{Patterns}d. We will check for the percolation between
fully filled $2 \times 2$ plaquettes or those with only one empty corner, see Fig.~\ref{Patterns}d.  We define the plaquette occupation number on the sites of the plaquette lattice as:
\begin{equation}
\label{nPl}
  \tilde n_{i,t}=\vartheta \Big(\sum_{\mathrm{plaq}} n_{i,t}-3 \Big)~,
\end{equation}
where the sum runs over 4 sites of the plaquette, and $\vartheta(x)$ is the Heaviside step function defined such that  $\vartheta(x \geq 0)=1$ and $\vartheta(x < 0)=0$.
We will be looking for the phase with percolation between active plaquettes connected by nn bonds ($PL$ phase).

\begin{figure*}[h]
\begin{subfigure}{0.4\textwidth}
    \includegraphics[width=7.2cm]{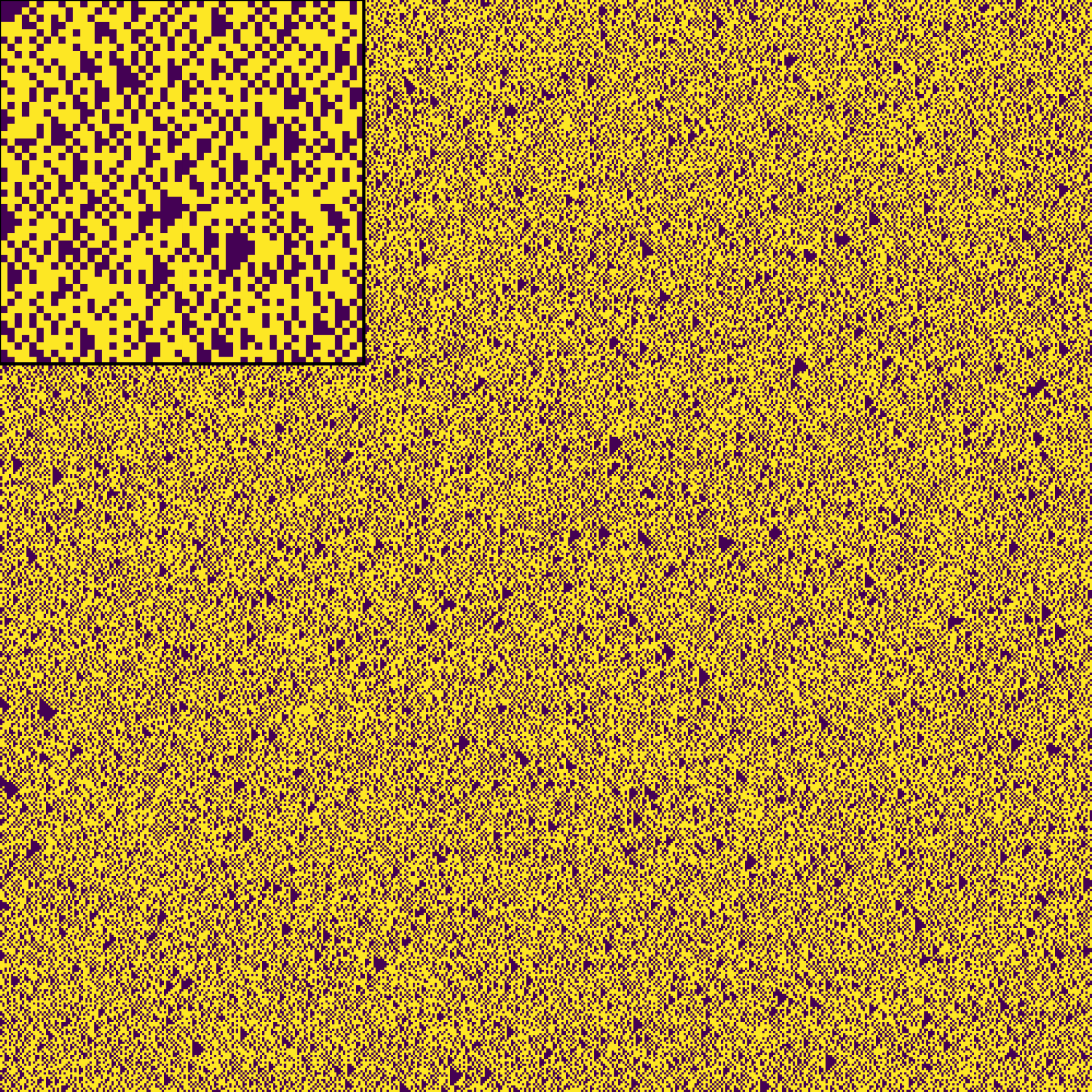}
    \caption{}
    \label{fig:1a}
  \end{subfigure}%
    \hspace{0.03\textwidth}
\begin{subfigure}{0.4\textwidth}
    \includegraphics[width=7.2cm]{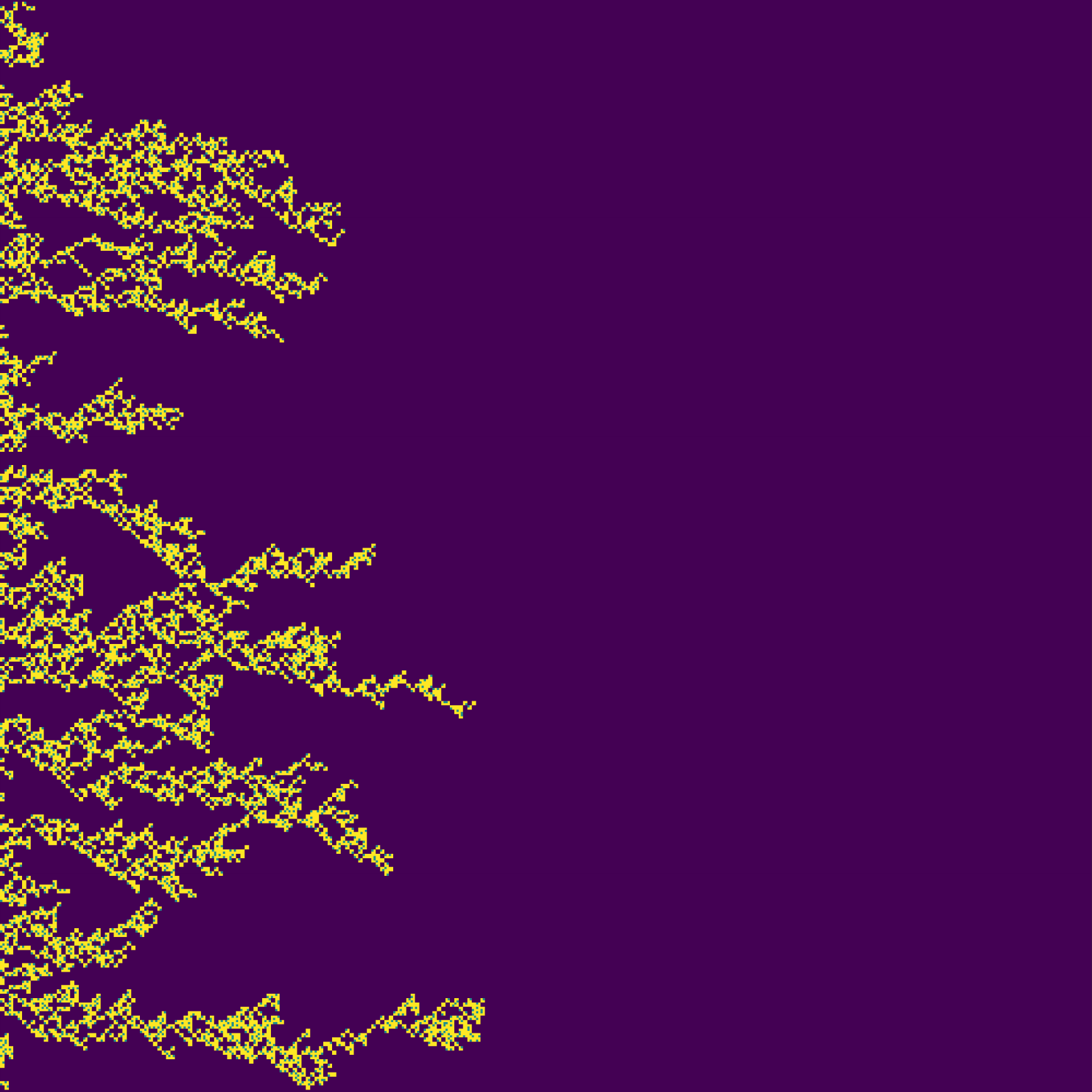}
    \caption{}
    \label{fig:1b}
  \end{subfigure}%

\begin{subfigure}{0.4\textwidth}
\includegraphics[width=7.2cm] {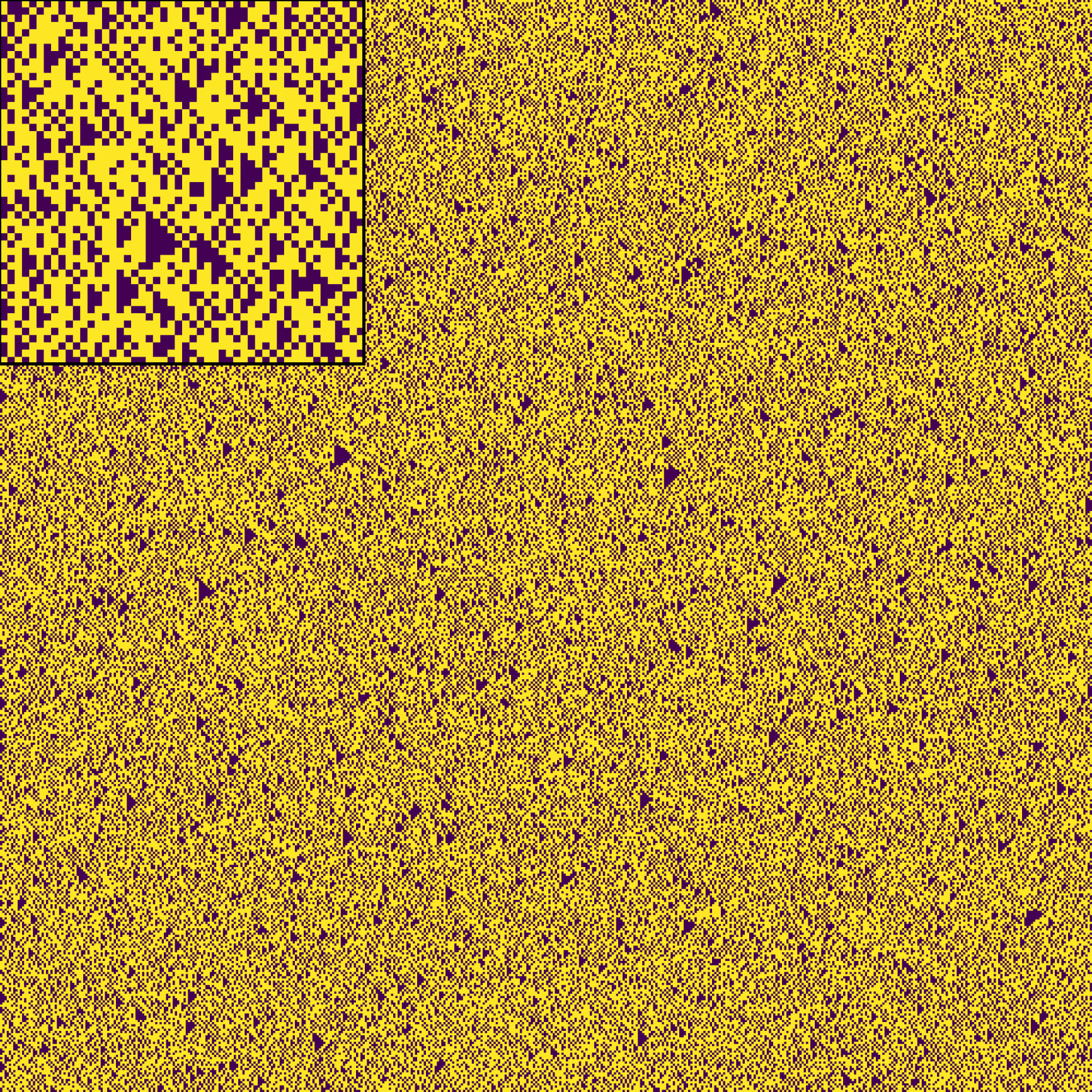}
    \caption{} \label{fig:1c}
  \end{subfigure}%
    \hspace{0.03\textwidth}
\begin{subfigure}{0.4\textwidth}
    \includegraphics[width=7.2cm]{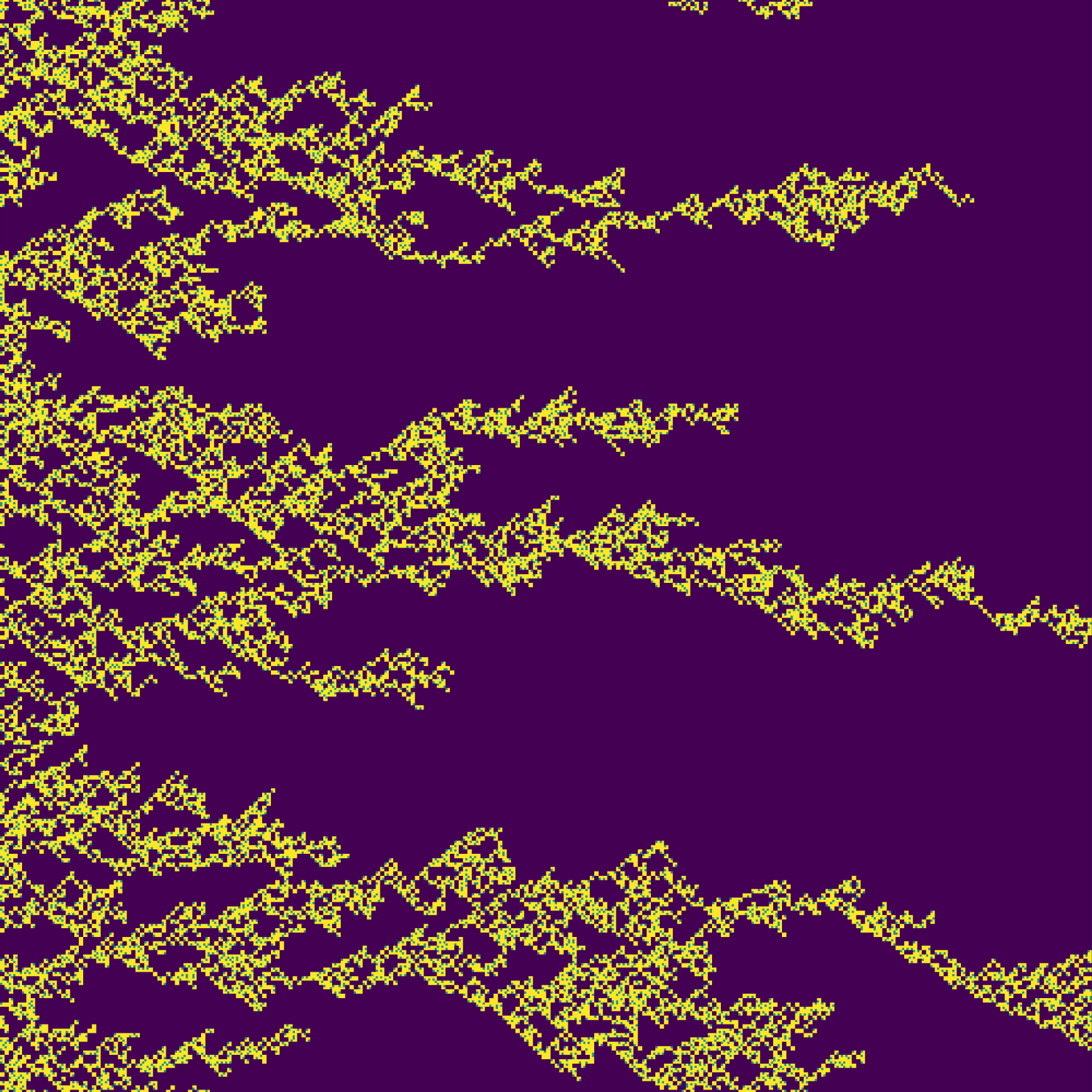}
    \caption{}
    \label{fig:1d}
  \end{subfigure}%
\caption{$PL$-patterns: (a,c) MC data ($N=500$, $T=500$) for two configurations of the percolating phase at $q=0.9$; (a) $p=0.49$, (c) $p=0.51$; critical point of the transition into $PL$ phase $p_c(q) \approx 0.498$; ochre/purple dots correspond to the filled/empty sites of the original lattice. Zoomed fragments of raw data shown in insets. (b,d) The connected plaquette patterns constructed from the data shown in (a,c); (b) $p<p_c(q)$, plaquette pattern is not percolative; (d) $p>p_c(q)$, the system is in the percolating $PL$ phase; yellow dots correspond to connected plaquettes; the dark purple background corresponds to the sites without plaquettes or to disconnected plaquettes without ancestors. }
\label{fig:PL_pattern_example}
\end{figure*}

Following the procedure used for other phases, we detect the $PL$ phase where it was expected, its location is shown in Fig.~\ref{PDiag}. Examples of the patterns and scaling analysis near transition are given in Fig.~\ref{fig:PL_pattern_example} and in Fig.~\ref{PLSC}, Appendix \ref{AppA}.

%
%%%%%%%%%%%%%%%%%%%%%%%%%%%%%%%%%%%%%%%%%%%%%%%%%%%%%%%%%%%%%%%%%%%%%%%%%%%%
%%%%%%%%%%%%%%%%%%%%%%%%%%%%%%%%%%%%%%%%%%%%%%%%%%%%%%%%%%%%%%%%%%%%%%%%%%%%
%%%%%%%%%%%%%%%%%%%%%%%%%%%%%%%%%%%%%%%%%%%%%%%%%%%%%%%%%%%%%%%%%%%%%%%%%%%
%%%%%%%%%%%%%%%%%%%%%%%%%%%%%%%%%%%%%%%%%%%%%%%%%%%%%%%%%%%%%%%%%%%%%%%%%%%
%

The results given in Table~\ref{Tab:PLtr}, Appendix \ref{AppA}
confirm the DP universality class of the transition into the $PL$ phase.

%
%
%xxxxxxxxxxxxxxxxxxxxxxxxxxxxxxxxxxxxxxxxxxxxxxxxxxxxxxxxxxxxxxxxxxxxxxxxxxxxxx
%xxxxxxxxxxxxxxxxxxxxxxxxxxxxxxxxxxxxxxxxxxxxxxxxxxxxxxxxxxxxxxxxxxxxxxxxxxxxxx
%xxxxxxxxxxxxxxxxxxxxxxxxxxxxxxxxxxxxxxxxxxxxxxxxxxxxxxxxxxxxxxxxxxxxxxxxxxxxxx
%
\section{Conclusion and Discussion}\label{Concl}
%xxxxxxxxxxxxxxxxxxxxxxxxxxxxxxxxxxxxxxxxxxxxxxxxxxxxxxxxxxxxxxxxxxxxxxxxxxxxxx
%xxxxxxxxxxxxxxxxxxxxxxxxxxxxxxxxxxxxxxxxxxxxxxxxxxxxxxxxxxxxxxxxxxxxxxxxxxxxxx
%xxxxxxxxxxxxxxxxxxxxxxxxxxxxxxxxxxxxxxxxxxxxxxxxxxxxxxxxxxxxxxxxxxxxxxxxxxxxxx
%
%

\begin{figure}[h]
%\vspace{-2em}
\centering{\includegraphics[width=8.5cm ]{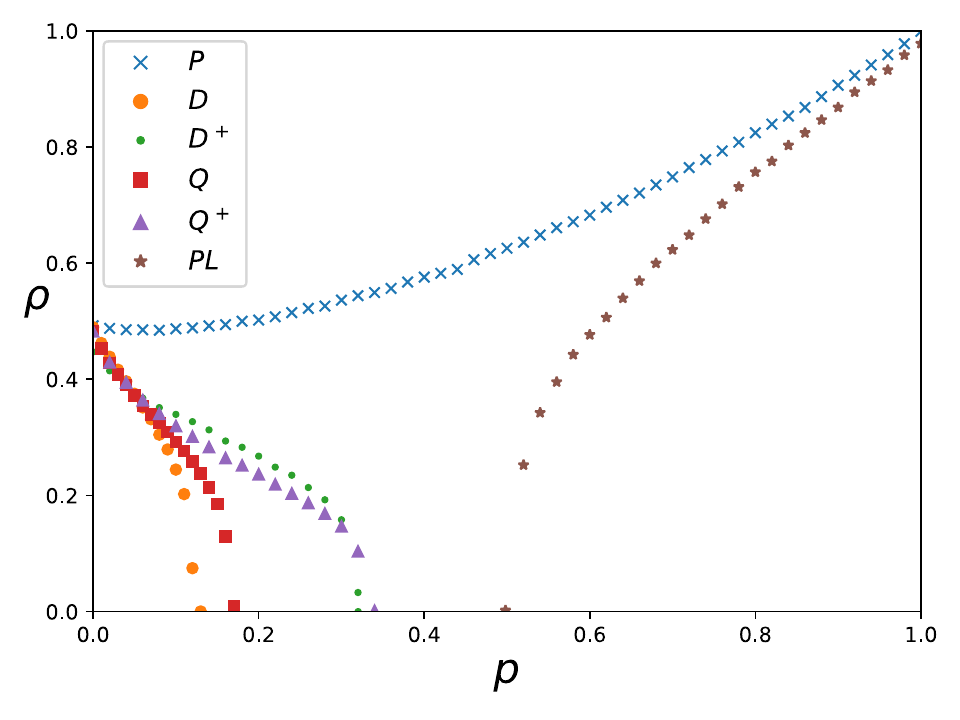}}
\caption{Hierarchy of percolation patterns: order parameters calculated along the line $q=0.9$ (within the active/percolating phase) for different values of $p$.}
\label{OPs}
%\vspace{-1em}
\end{figure}

The phase diagram of a one-dimensional kinetic contact process with parallel update is established using the Monte Carlo simulations and finite-size scaling, see Fig.~\ref{PDiag}.
The structure of the hidden percolative patterns (order parameters) emerging through transitions in the active phase and the nature of those transitions are revealed.
Our results corroborate the conjecture that in general the active (percolating) phases possess the hierarchical structure (tower of percolation patterns), where more complicated patterns emerge on the top of coexistent patterns of lesser complexity, see Fig.~\ref{OPs}. Plethora of different patterns emerge via cascades of continuous phase transitions. We detect five phases with distinct patterns of percolation within the active phase of the model. From the results visualized in Figs.~\ref{PDiag} and \ref{OPs} we infer the following hierarchy in the active phase: at $q=0.9$ and $p \sim 0.4$ the model is just percolating, it sits in the ``trivial" $P$-phase (see the ``flat" non-vanishing $P$-order parameter in the whole region $p \in [0,1]$ shown in Fig.~\ref{OPs}). Moving from there along $q=0.9$ towards $p \to 0$, the tower of coexistent percolative patterns builds up:
\begin{eqnarray}
\label{Hleft}
  &q&=0.9,p<0.4,~~ p  \rightarrowtail 0: \nonumber \\
  &P& \rightarrow (P+Q^+) \rightarrow (P+Q^+ + D^+) \rightarrow  \\
   &(&P+Q^+ + D^+ +Q) \rightarrow (P+Q^+ +D^+ +Q+D)~,    \nonumber
\end{eqnarray}
as we cross the  corresponding critical lines.
The percolarive landscape is less rich in the opposite direction:
\begin{equation}
\label{Hright}
 q=0.9,p>0.5,~~ p  \rightarrowtail 1:~P \rightarrow (P+PL)~.
\end{equation}
The hierarchy \eqref{Hleft} and \eqref{Hright} appears to be a judicious set
of phases to account for distinct local order easily seen in the percolating patterns on the parametric $(p,q)$ plane of the model.
In line with the general conjecture about the infinite cascade of percolative transitions, we could have searched for
the $4 \times 4$ plaquettes, and probably even more sophisticated and/or bigger patterns. Keeping in mind applications of these ideas to the comparative
quantitative analysis of connectivity in different systems, especially in networks with the use of machine-learning codes (neural networks),
we restricted our analysis to the judicious set of quite simple and visually distinguished patterns.

%
%
%
%%%%%%%%%%%%%%%%%%%%%%%%%%%%%%%%%%%%%%%%%%%%%%%%%%%%%%%%%%%%%%%%%%%%%%%%%%%%
%%%%%%%%%%%%%%%%%%%%%%%%%%%%%%%%%%%%%%%%%%%%%%%%%%%%%%%%%%%%%%%%%%%%%%%%%%%%
%%%%%%%%%%%%%%%%%%%%%%%%%%%%%%%%%%%%%%%%%%%%%%%%%%%%%%%%%%%%%%%%%%%%%%%%%%%%
%%%%%%%%%%%%%%%%%%%%%%%%%%%%%%%%%%%%%%%%%%%%%%%%%%%%%%%%%%%%%%%%%%%%%%%%%%%%
%
%
%

All transitions on the phase diagram belong to the DP universality class,  as confirmed by the finite-size scaling analysis. This seems to be at odds with the Janssen-Grassberger (JG) conjecture \cite{Janssen:1981,Grassberger:1982}.
The universality classes of various nonequilibrium lattice models are well-classified \cite{Odor:2004}. In particular, according to the JG conjecture, a model belongs to the DP universality class if it satisfies the following conditions: (i) transition is continuous between a fluctuating active phase and a unique absorbing state; (ii) the order parameter is positive and one-component; (iii) interactions are short-ranged; (iv) no quenched disorder or additional symmetries are present \cite{Hinrichsen:2006}.
Apparently the requirement (i) needs to be clarified to accommodate the case of multiple percolating phases. For each active (percolating) phase analyzed, the requirements (ii)-(iv) are satisfied, although the active phase is not unique. The absorbing (empty) state (vacuum $|0 \rangle_{\sharp}$) is unique for each active phase, but those empty states (vacua) are defined with respect to a particular pattern (see
Figs.~\ref{fig:dipole_pattern_example},\ref{fig:quadrupole_pattern_example},\ref{fig:dipole+_pattern_example},\ref{fig:quadrupole+_pattern_example}
and comments to those figures), so they are distinct states:
\begin{equation}
\label{vacua}
  |0 \rangle_{P} \neq |0 \rangle_{Q},~\mathrm{etc},~\forall ~P,Q,Q^+,D,D^+,PL~.
\end{equation}
The present results suggest an extension of the JG conjecture. It holds if the first condition above is modified as follows:\\
($\mathrm{i}^\prime$) transition is continuous between a fluctuating active (percolating) phase $|\mathrm{Ac} \rangle_{i}$ and its own unique absorbing (empty) state $|0 \rangle_{i}$; the couple
active-empty state is not unique in general:
$\{|\mathrm{Ac} \rangle_{i},|0 \rangle_{i}\}$, $i=1,2,\ldots~~$.

Thus far the results on the tower of percolating patterns are deduced from the numerical analysis of simulations. To get a better insight on the origin of the cascades, a promising direction is to relate these transitions to zeros of the non-equilibrium stationary partition functions. Yang and Lee \cite{YangLee:1952,*LeeYang:1952} pioneered this rigorous statistical-mechanical approach to study phase transitions. The theory is applicable whatever are the nature of order parameter or symmetry breaking. It can be extended for the open systems and models out of equilibrium, see, e.g., \cite{Arndt:2000,Blythe:2002,Hinrichsen:2002,Bena:2005,Heyl:2018,Ueda:2022}, and more references there.
In several equilibrium models \cite{Chitov:2017PRE,*Chitov:2021,*Chitov:2022DL} one can relate the Lee-Yang zeros to the cascades of transitions and also to the spectra of the transfer matrices. So, the goal for the future work is to relate zeros of generalized partition functions of the stationary states to the spectra of transfer matrices deduced from the master equation of the kinetic replication model \eqref{ModDef}.

On the purely computational side, it is very appealing to work out some machinery for the percolation pattern recognition in the raw MC data. The neural networks/machine learning (for a review, see e.g., \cite{Kapitan:2023} and more references there) seems to be a very effective tool for the problems of percolation  \cite{Cheng:2021,Shen:2021,*Shen:2022}
and even for the implementation of the Lee-Yang approach \cite{Flindt:2023}. Progress has been made recently in applying neural networks to detect the hidden percolation patterns of the model \eqref{ModDef} and to quantitatively reproduce its phase diagram shown in Fig.~\ref{PDiag}, to be reported elsewhere.

The networks is the field where the above ideas and approaches can be most naturally applied and advanced. It is useful to draw more analogies between the current results and the well-known results in the theory of networks (graphs). In the standard treatment of  percolation or cellular automata, the main task is to detect transition into the percolation (active) phase which is signalled by appearance of the connected cluster running through the whole system. A possibility that this phase has the internal complexity, structure or hierarchy, is usually beyond the scope of the standard approach. On the other hand, in the network (graph) theory, studies of modularity and structural complexity of a connected graph (giant component) is a very active field of research \cite{Newman:2010,Fortunato:2010,Sporns:2010,*Sporns:2012,Barabasi:2016,Munoz:2018}. Modules (communities) in the graph theory are defined as groups of nodes which share greater number of mutual connection within each module (community) and fewer connections between modules \cite{Newman:2010}. The percolation patterns shown in Fig.~\ref{Patterns} qualify for the definition of modules (communities) on the regular graph (lattice).
Practically all real-world networks, ranging from the Internet, logistics, transport to the human connectome, demonstrate the modular and hierarchical organization
\cite{Fortunato:2010,Sporns:2010,*Sporns:2012,Barabasi:2016,Munoz:2018}. So, the main goal of the present study -- the search of the hidden structure of the percolating phase -- is analogous to the detection of the modules in the networks (graphs). The model \eqref{ModDef} generates quite diverse percolating phase on the parametric plane $(p,q)$. In the central part of the diagram where only the simple $P$ phase is present, the percolating landscape can be compared to a giant component of the Erd\"{o}s-R\'{e}nyi  random graph which does not have any modularity \cite{Fortunato:2010}. The connectivity and modularity of a network (empirical or model-generated) can be characterized, e.g.,  by the degree distribution of nodes or betweenness centrality \cite{Newman:2010}. However those characteristics are of the cross-over type,
while the percolation patterns revealed in this work, clearly demonstrate critical properties of the order parameter. \\
\textit{(i):} A substantial extension of the present results would be to quantitatively classify the organization of a network (that is its functionality) by its tower of the patterns of connectivity.\\
\textit{(ii):} A closely related very important practical problem in generic networks (e.g., communications, logistics, traffic, etc) is their resilience. The power-law networks are very robust with respect to random failures, but are vulnerable to the targeted attacks removing primarily highly connected nodes (hubs) \cite{Newman:2010,Barabasi:2016}. The latter implies that the resilience is directly related to the hierarchical modular organization of a given network. The goal then is to quantify the resilience of a network in terms of the potential damage to its hierarchy of the connectivity patterns \textit{(i)}.  If applied to brain networks, the functionality for instance of a connectome, classified by its tower of judicious patterns of connectivity, is expected to be related to the different states of mind, like norm, disease, etc. Those are directions of future work.

%
%xxxxxxxxxxxxxxxxxxxxxxxxxxxxxxxxxxxxxxxxxxxxxxxxxxxxxxxxxxxxxxxxxxxxxxxxxxxxxx
\begin{acknowledgments}
This research is supported by the grant \# 24-22-00075
(https://rscf.ru/project/24-22-00075/)
from the Russian Science Foundation.
\end{acknowledgments}
%xxxxxxxxxxxxxxxxxxxxxxxxxxxxxxxxxxxxxxxxxxxxxxxxxxxxxxxxxxxxxxxxxxxxxxxxxxxxxx

\bibliography{C:/Users/gchitov/Documents/Papers/BibRef/BibTexCombined.bib}
%\bibliography{BibTexCombined.bib}
%\bibliography{BibTexFile.bib}

%\clearpage

%%%%%%%%%%%%%%%%%%%%%%%%%%%%%%%%%%%%%%%%%%%%%%%%%%%%%%%%%%%%%%%%%%%%%%%%%%%%%%
%%%%%%%%%%%%%%%%%%%%%%%%%%%%%%%%%%%%%%%%%%%%%%%%%%%%%%%%%%%%%%%%%%%%%%%%%%%%%%
\begin{appendix}
\section{Scaling data for $Q$ and $PL$ phases}\label{AppA}
%%%%%%%%%%%%%%%%%%%%%%%%%%%%%%%%%%%%%%%%%%%%%%%%%%%%%%%%%%%%%%%%%%%%%%%%%%%%%%
%%%%%%%%%%%%%%%%%%%%%%%%%%%%%%%%%%%%%%%%%%%%%%%%%%%%%%%%%%%%%%%%%%%%%%%%%%%%%%
%
%
To unload the main text, we collect in this Appendix the tables with
the critical parameters $q_{c}$ and critical indices for the $Q$ and $PL$ phases,
along with representatives examples of the finite-size scaling analysis, discussed in the main text.

%
%%%%%%%%%%%%%%%%%%%%%%%%%%%%%%%%%%%%%%%%%%%%%%%%%%%%%%%%%%%%%%%%%%%%%%%%%%%%
%%%%%%%%%%%%%%%%%%%%%%%%%%%%%%%%%%%%%%%%%%%%%%%%%%%%%%%%%%%%%%%%%%%%%%%%%%%%
\begin{table}[h]
\centering
\caption{Critical points $q_{c}$ and critical indices for the transition into $Q$ phase for several values of $p$. Scaling analysis presented in Fig.~\ref{QSC} yields parameters shown in bold. }
{%
\begin{tabular}{|l|l|l|l|l|l|l|}
\hline
$p$ & $q_c$ & $\alpha$ & $\nu_{||}$ & z & $\beta = \alpha\nu_{||}$ & $\nu_{\perp} = \nu_{||} / z $ \\ \hline
0.0000 & 0.6564 & 0.1590 & 1.72 & 1.57 & 0.27(3) & 1.09(5) \\ \hline
\textbf{0.1000} & \textbf{0.7656} & \textbf{0.1595} & \textbf{1.72} & \textbf{1.56} & \textbf{0.27(4)} & \textbf{1.10(2)} \\ \hline
0.1500 & 0.8557 & 0.1590 & 1.72 & 1.58 & 0.27(3) & 1.08(8) \\ \hline
0.1700 & 0.9161 & 0.1590 & 1.72 & 1.55 & 0.27(3) & 1.10(9) \\ \hline
0.1800 & 1.0000 & 0.1590 & 1.72 & 1.58 & 0.27(3) & 1.08(8) \\ \hline
\end{tabular}%
}
\label{Tab:Qtr}
\end{table}
%%%%%%%%%%%%%%%%%%%%%%%%%%%%%%%%%%%%%%%%%%%%%%%%%%%%%%%%%%%%%%%%%%%%%%%%%%%%
%

%
%
%
%%%%%%%%%%%%%%%%%%%%%%%%%%%%%%%%%%%%%%%%%%%%%%%%%%%%%%%%%%%%%%%%%%%%%%%%%%%%
%%%%%%%%%%%%%%%%%%%%%%%%%%%%%%%%%%%%%%%%%%%%%%%%%%%%%%%%%%%%%%%%%%%%%%%%%%%%
\begin{figure*}[h]
\begin{subfigure}{0.4\textwidth}
    \includegraphics[width=7.2cm]{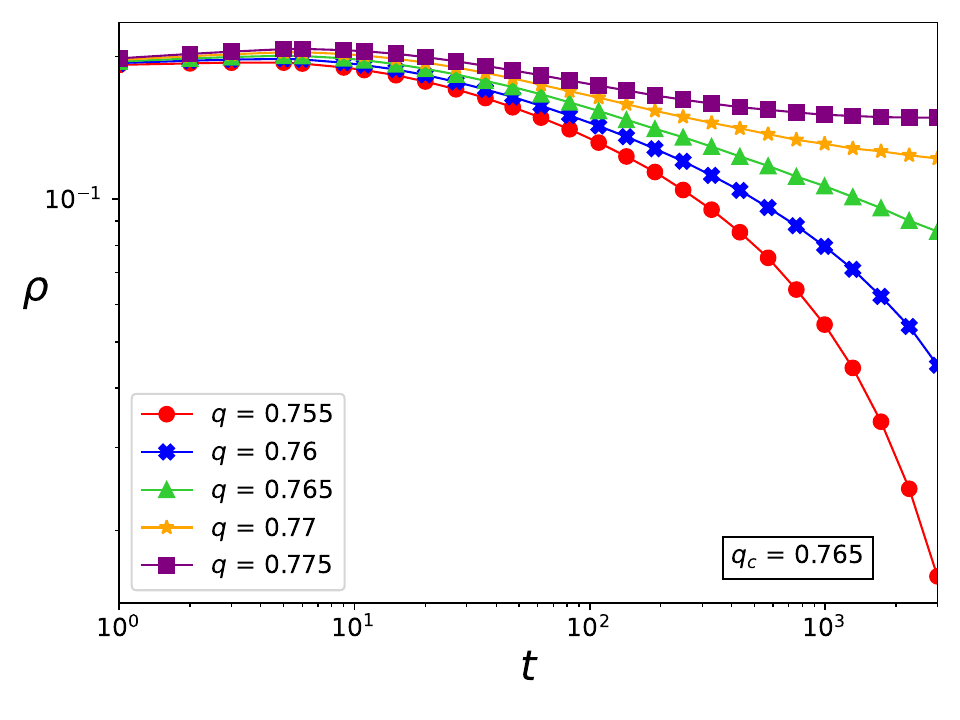}
    \caption{} \label{fig:2a}
  \end{subfigure}%
\begin{subfigure}{0.4\textwidth}
    \includegraphics[width=7.2cm]{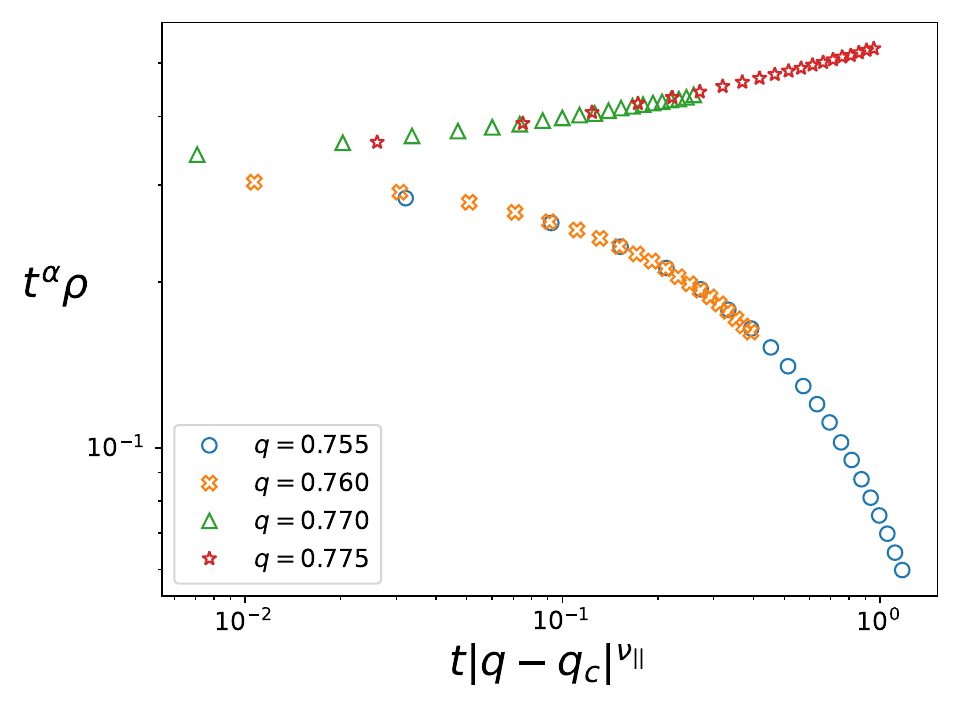}
    \caption{} \label{fig:2b}
  \end{subfigure}%

\begin{subfigure}{0.4\textwidth}
\hspace*{-7mm}
\includegraphics[width=6.7cm]
    {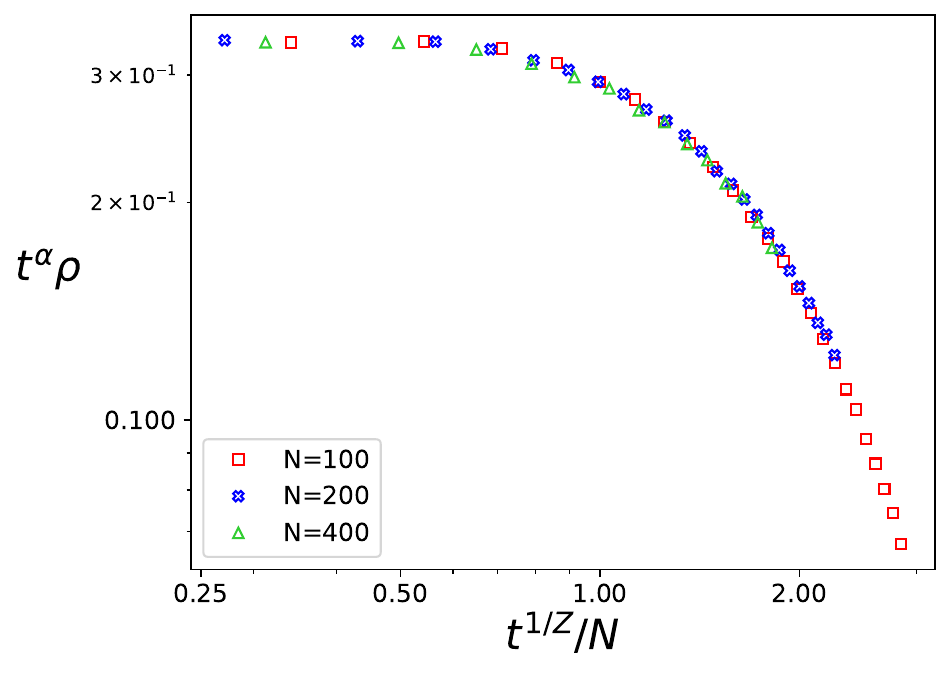}
    \caption{} \label{fig:2c}
  \end{subfigure}%
\begin{subfigure}{0.4\textwidth}
    \includegraphics[width=7.2cm]{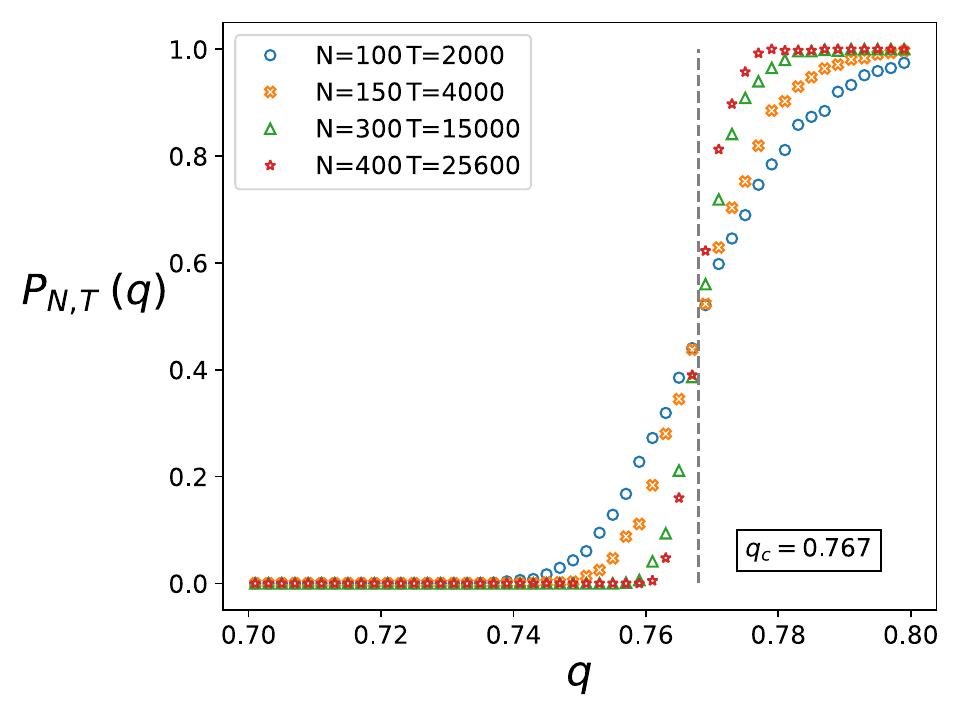}
    \caption{} \label{fig:2d}
  \end{subfigure}%
\caption{$Q$-phase: (a) Density of the connected (nn) quadrupoles $\rho(t)$ for $p=0.1$ and series of $q$ near $q_c \approx 0.7656$. The system size is $N=3000$ and $T=100000$. The middle line corresponds to the critical slowing down $\rho(t) \propto t^{- \alpha}$ with $\alpha \approx 0.1595$.  (b) Collapse of the curves from (a) onto a single scaling function. Fitting gives the values of $q_c$ and $\nu_\parallel$. (c) Collapse of the order parameter relaxation $t^{\alpha}\rho(t)$ at the critical point  for different sizes
($N=100;~200;~400$) yields the critical index $z$. (d) Fraction of connected quadrupoles $P_{N,T}(q)$ for different sizes $N \times T$, all curves intersect at  $q_c \approx 0.767$. The critical values of $q_c$ obtained from (a), (b) and (d) agree within $\sim 0.01 \% $}
\label{QSC}
\end{figure*}
%%%%%%%%%%%%%%%%%%%%%%%%%%%%%%%%%%%%%%%%%%%%%%%%%%%%%%%%%%%%%%%%%%%%%%%%%%%%
%%%%%%%%%%%%%%%%%%%%%%%%%%%%%%%%%%%%%%%%%%%%%%%%%%%%%%%%%%%%%%%%%%%%%%%%%%%%%%%%%%%%%%%%%%%%%%%%%%%%%%%%%%%%%%%%%%%%%%%%%%%%%%%%%%%%%%
%%%%%%%%%%%%%%%%%%%%%%%%%%%%%%%%%%%%%%%%%%%%%%%%%%%%%%%%%%%%%%%%%%%%%%%%%%%%

%
%%%%%%%%%%%%%%%%%%%%%%%%%%%%%%%%%%%%%%%%%%%%%%%%%%%%%%%%%%%%%%%%%%%%%%%%%%%%
%%%%%%%%%%%%%%%%%%%%%%%%%%%%%%%%%%%%%%%%%%%%%%%%%%%%%%%%%%%%%%%%%%%%%%%%%%%%
\begin{table}[h]
\centering
\caption{Critical points $q_{c}$ and critical indices for the transition into $PL$ phase for several values of $p$. Scaling analysis presented in Fig.~\ref{PLSC} yields parameters shown in bold. }
{%
\begin{tabular}{|l|l|l|l|l|l|l|}
\hline
$p$ & $q_c$ & $\alpha$ & $\nu_{||}$ & z & $\beta = \alpha\nu_{||}$ & $\nu_{\perp} = \nu_{||} / z $ \\ \hline
0.4900 & 1.0000 & 0.1590 & 1.72 & 1.58 & 0.27(3) & 1.08(8) \\ \hline
0.6000 & 0.6256 & 0.1590 & 1.72 & 1.54 & 0.27(3) & 1.11(6) \\ \hline
\textbf{0.7000} & \textbf{0.4854} & \textbf{0.1590} & \textbf{1.72} & \textbf{1.55} & \textbf{0.27(3)} & \textbf{1.10(9)} \\ \hline
0.8000 & 0.3395 & 0.1590 & 1.72 & 1.58 & 0.27(3) & 1.08(8) \\ \hline
0.9000 & 0.1865 & 0.1593 & 1.72 & 1.56 & 0.27(4) & 1.10(2) \\ \hline
\end{tabular}%
}
\label{Tab:PLtr}
\end{table}
%%%%%%%%%%%%%%%%%%%%%%%%%%%%%%%%%%%%%%%%%%%%%%%%%%%%%%%%%%%%%%%%%%%%%%%%%%%%
%%%%%%%%%%%%%%%%%%%%%%%%%%%%%%%%%%%%%%%%%%%%%%%%%%%%%%%%%%%%%%%%%%%%%%%%%%%%
%

%
%
%%%%%%%%%%%%%%%%%%%%%%%%%%%%%%%%%%%%%%%%%%%%%%%%%%%%%%%%%%%%%%%%%%%%%%%%%%%%
%%%%%%%%%%%%%%%%%%%%%%%%%%%%%%%%%%%%%%%%%%%%%%%%%%%%%%%%%%%%%%%%%%%%%%%%%%%%
\begin{figure*}[h]
\begin{subfigure}{0.4\textwidth}
    \includegraphics[width=7.2cm]{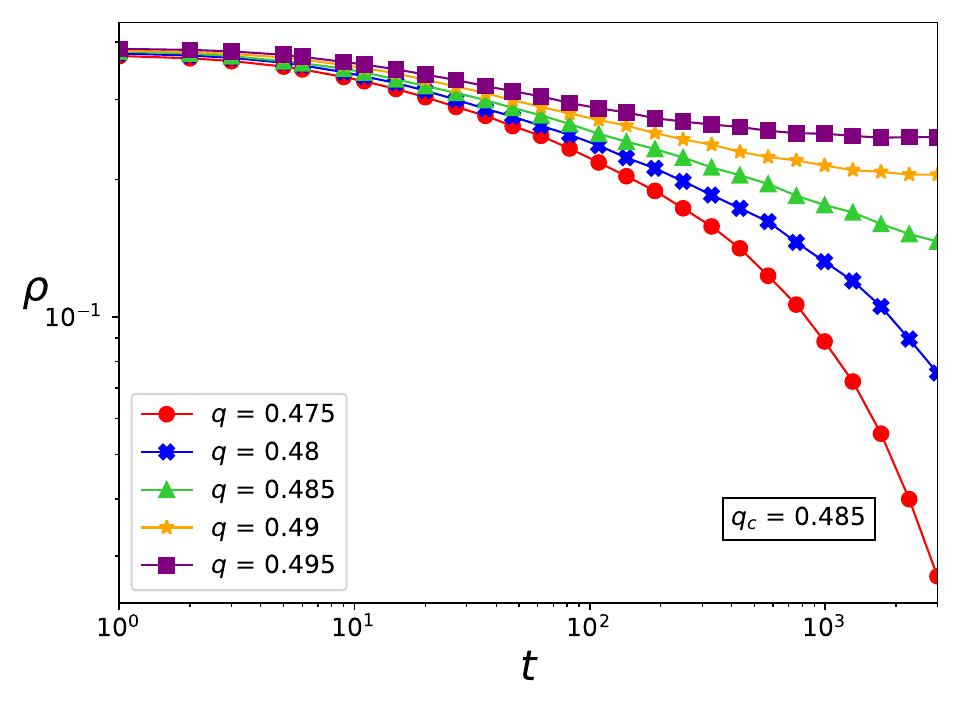}
    \caption{} \label{fig:4a}
  \end{subfigure}%
\begin{subfigure}{0.4\textwidth}
    \includegraphics[width=7.2cm]{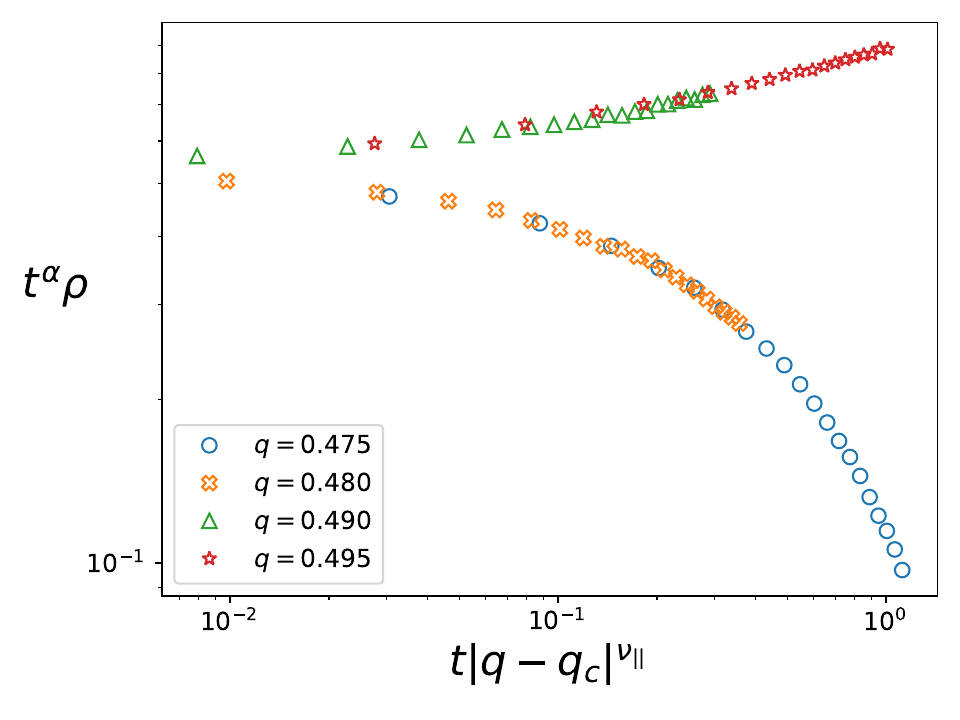}
    \caption{} \label{fig:4b}
  \end{subfigure}%

\begin{subfigure}{0.4\textwidth}
\hspace*{-7mm}
\includegraphics[width=6.7cm]
{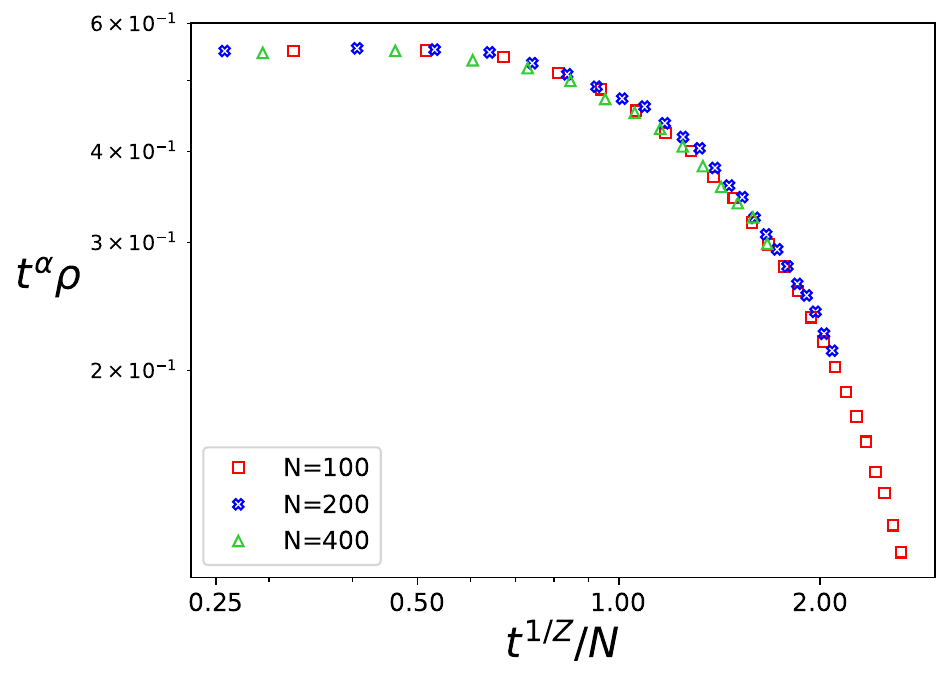}
    \caption{} \label{fig:4c}
  \end{subfigure}%
\begin{subfigure}{0.4\textwidth}
    \includegraphics[width=7.2cm]{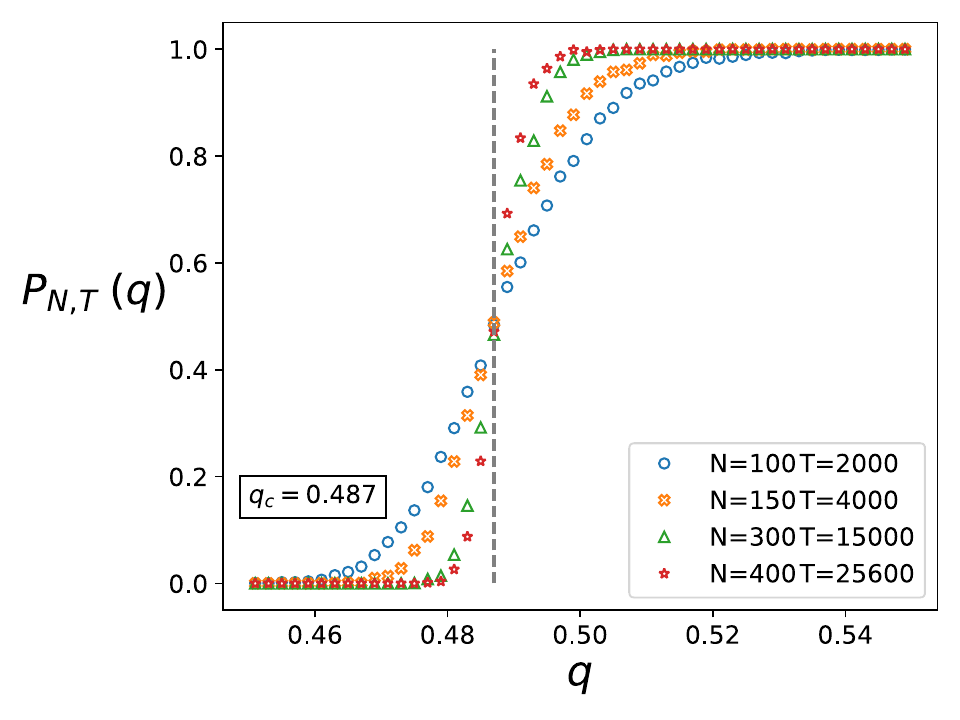}
    \caption{} \label{fig:4d}
  \end{subfigure}%
\caption{$PL$-phase: (a) Density of connected plaquettes $\rho(t)$ for $p=0.7$ and series of $q$ near $q_c \approx 0.4854$. The system size is $N=3000$ and $T=100000$. The middle line corresponds to the critical slowing down $\rho(t) \propto t^{- \alpha}$ with $\alpha \approx 0.1590$.  (b) Collapse of the curves from (a) onto a single scaling function. Fitting gives the values of $q_c$ and $\nu_\parallel$. (c) Collapse of the order parameter relaxation $t^{\alpha}\rho(t)$ at the critical point  for different sizes
 ($N=100;~200;~400$) yields the critical index $z$. (d) Fraction of connected plaquettes $P_{N,T}(q)$ for different sizes $N \times T$, all curves intersect
 at  $q_c \approx 0.487$. The critical values of $q_c$ obtained from (a), (b) and (d) agree within $\sim 0.01 \% $}
\label{PLSC}
\end{figure*}
%%%%%%%%%%%%%%%%%%%%%%%%%%%%%%%%%%%%%%%%%%%%%%%%%%%%%%%%%%%%%%%%%%%%%%%%%%%%
%%%%%%%%%%%%%%%%%%%%%%%%%%%%%%%%%%%%%%%%%%%%%%%%%%%%%%%%%%%%%%%%%%%%%%%%%%%%
%
%
%

%
%
%%%%%%%%%%%%%%%%%%%%%%%%%%%%%%%%%%%%%%%%%%%%%%%%%%%%%%%%%%%%%%%%%%%%%%%%%%%%%%
%%%%%%%%%%%%%%%%%%%%%%%%%%%%%%%%%%%%%%%%%%%%%%%%%%%%%%%%%%%%%%%%%%%%%%%%%%%%%%
\section{Numerical results for $D^+$ and $Q^+$ phases}\label{AppB}
%%%%%%%%%%%%%%%%%%%%%%%%%%%%%%%%%%%%%%%%%%%%%%%%%%%%%%%%%%%%%%%%%%%%%%%%%%%%%%
%%%%%%%%%%%%%%%%%%%%%%%%%%%%%%%%%%%%%%%%%%%%%%%%%%%%%%%%%%%%%%%%%%%%%%%%%%%%%%
%
%
%\begin{widetext}
The numerical results and analysis of $D^+$ and $Q^+$ phases presented in this Appendix. This material is pertinent to corroborate the consistency of overall results and conclusions.

\begin{figure*}[h]
\begin{subfigure}{0.4\textwidth}
    \includegraphics[width=7.2cm]{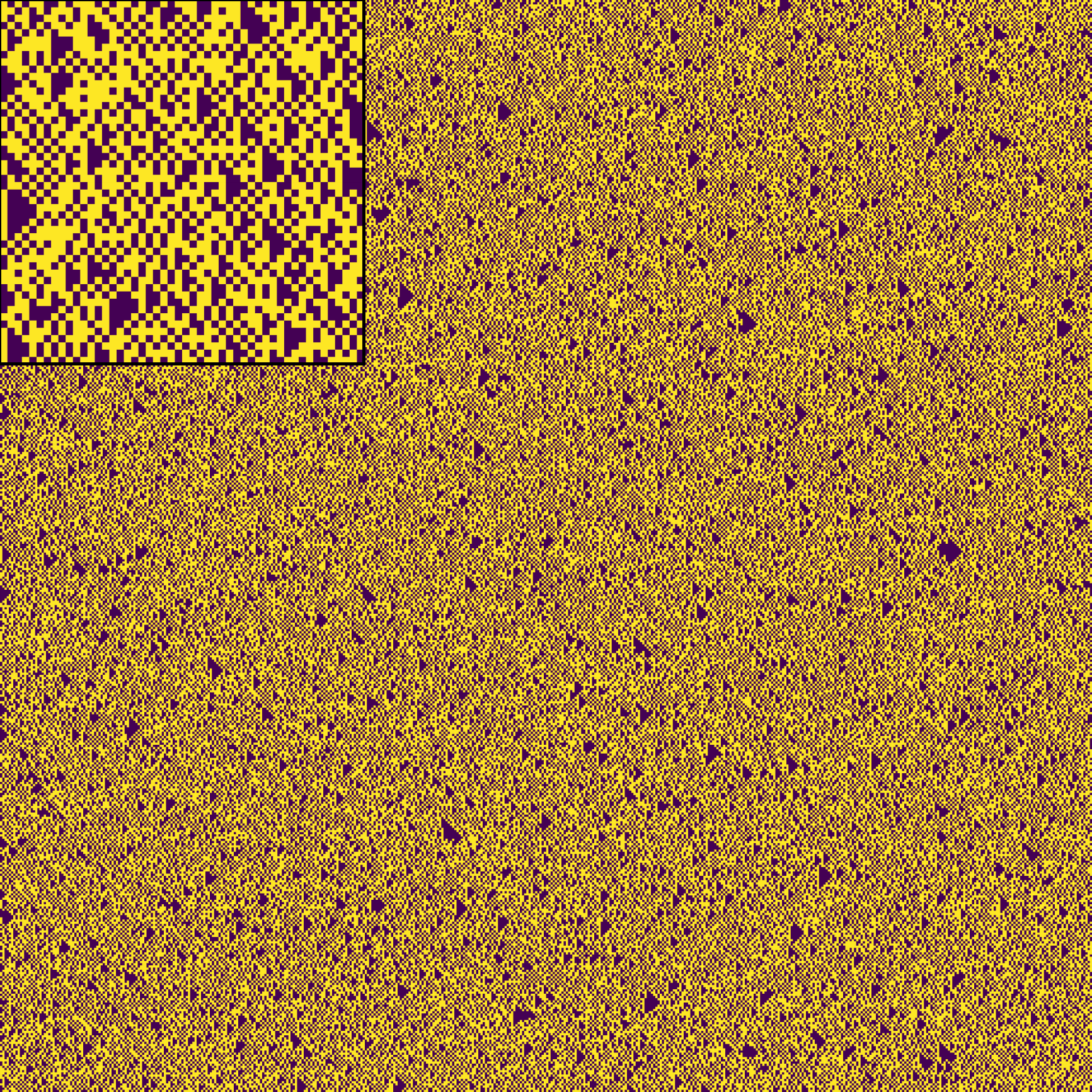}
    \caption{}
    \label{fig:1a}
  \end{subfigure}%
    \hspace{0.03\textwidth}
\begin{subfigure}{0.4\textwidth}
    \includegraphics[width=7.2cm]{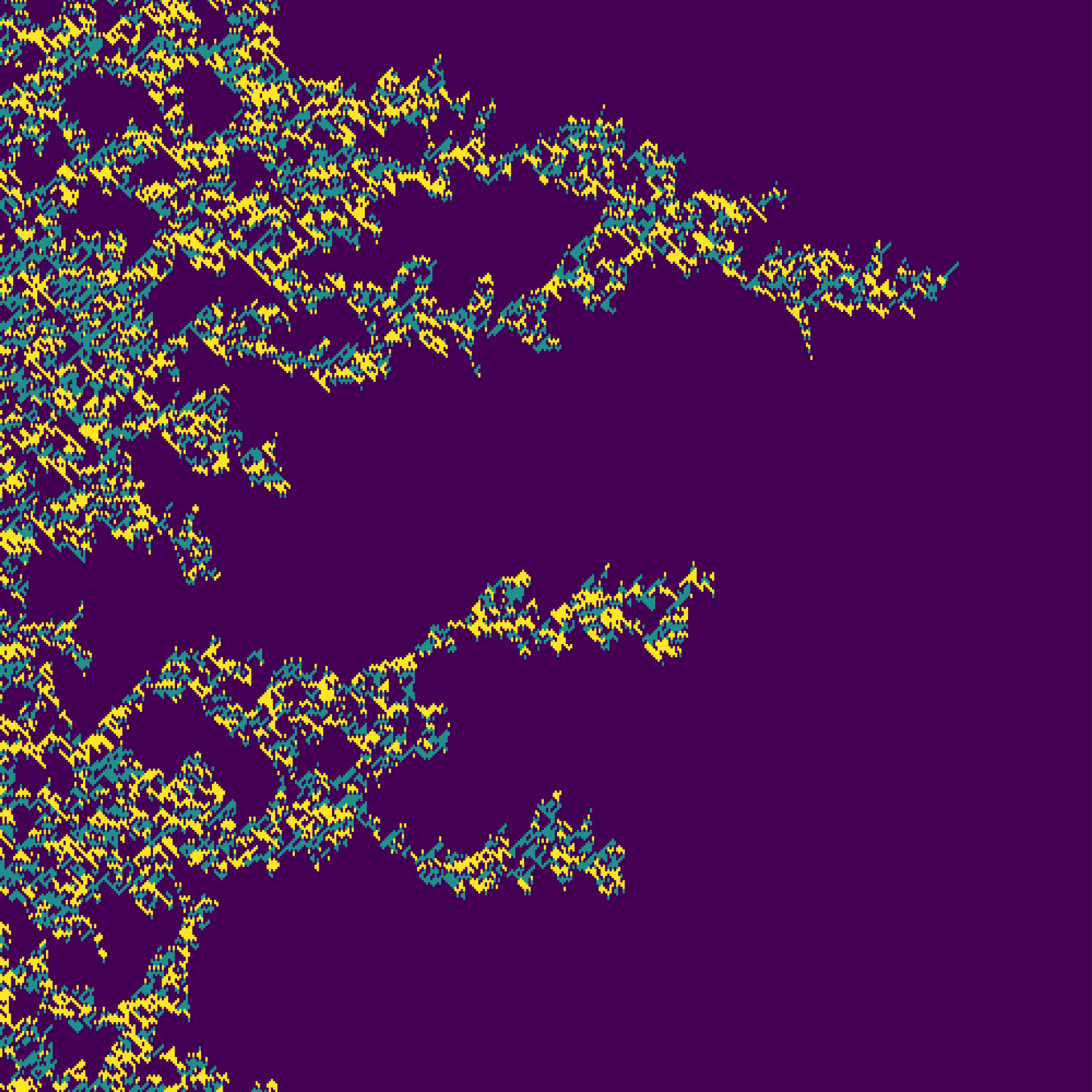}
    \caption{}
    \label{fig:1b}
  \end{subfigure}%

\begin{subfigure}{0.4\textwidth}
\includegraphics[width=7.2cm] {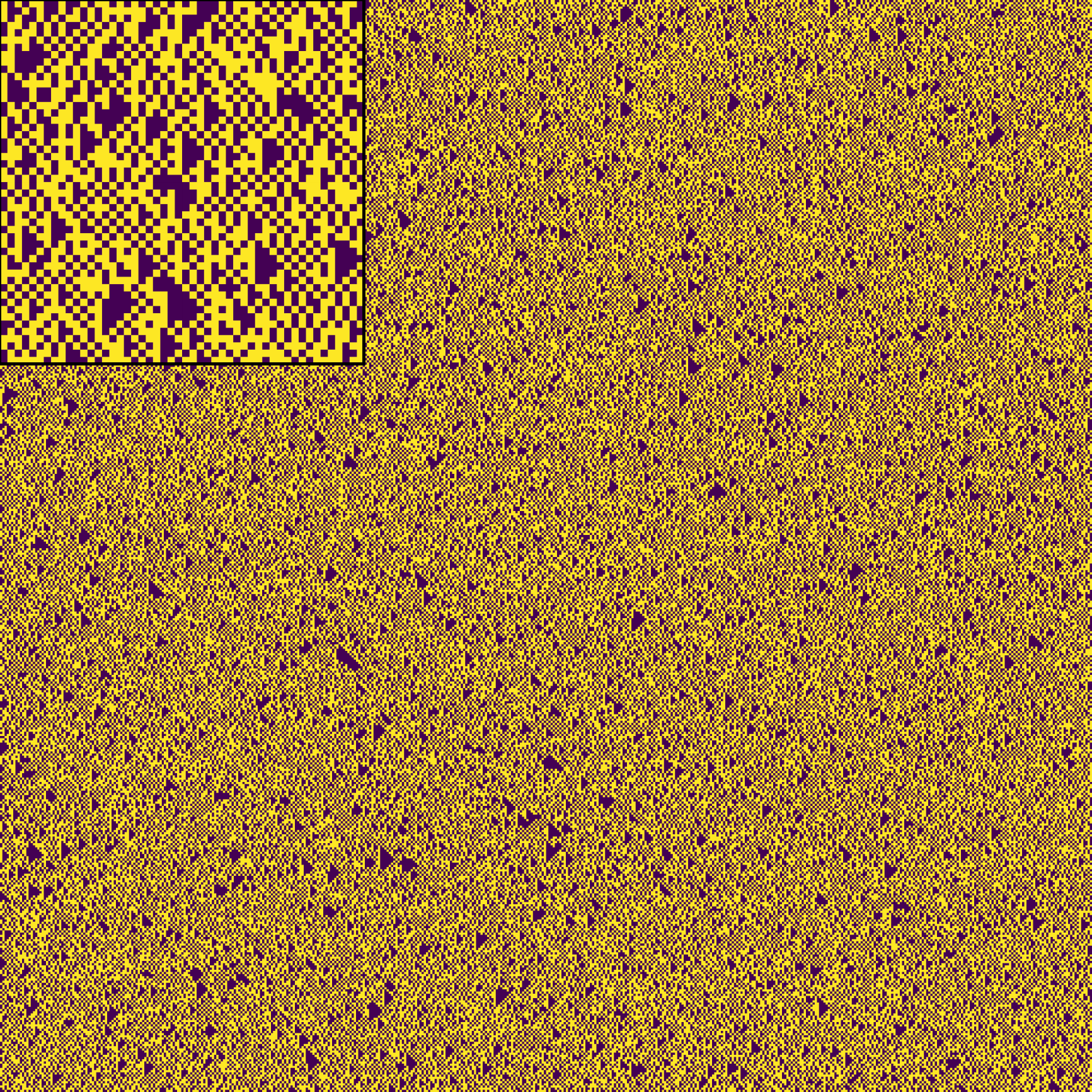}
    \caption{} \label{fig:1c}
  \end{subfigure}%
    \hspace{0.03\textwidth}
\begin{subfigure}{0.4\textwidth}
    \includegraphics[width=7.2cm]{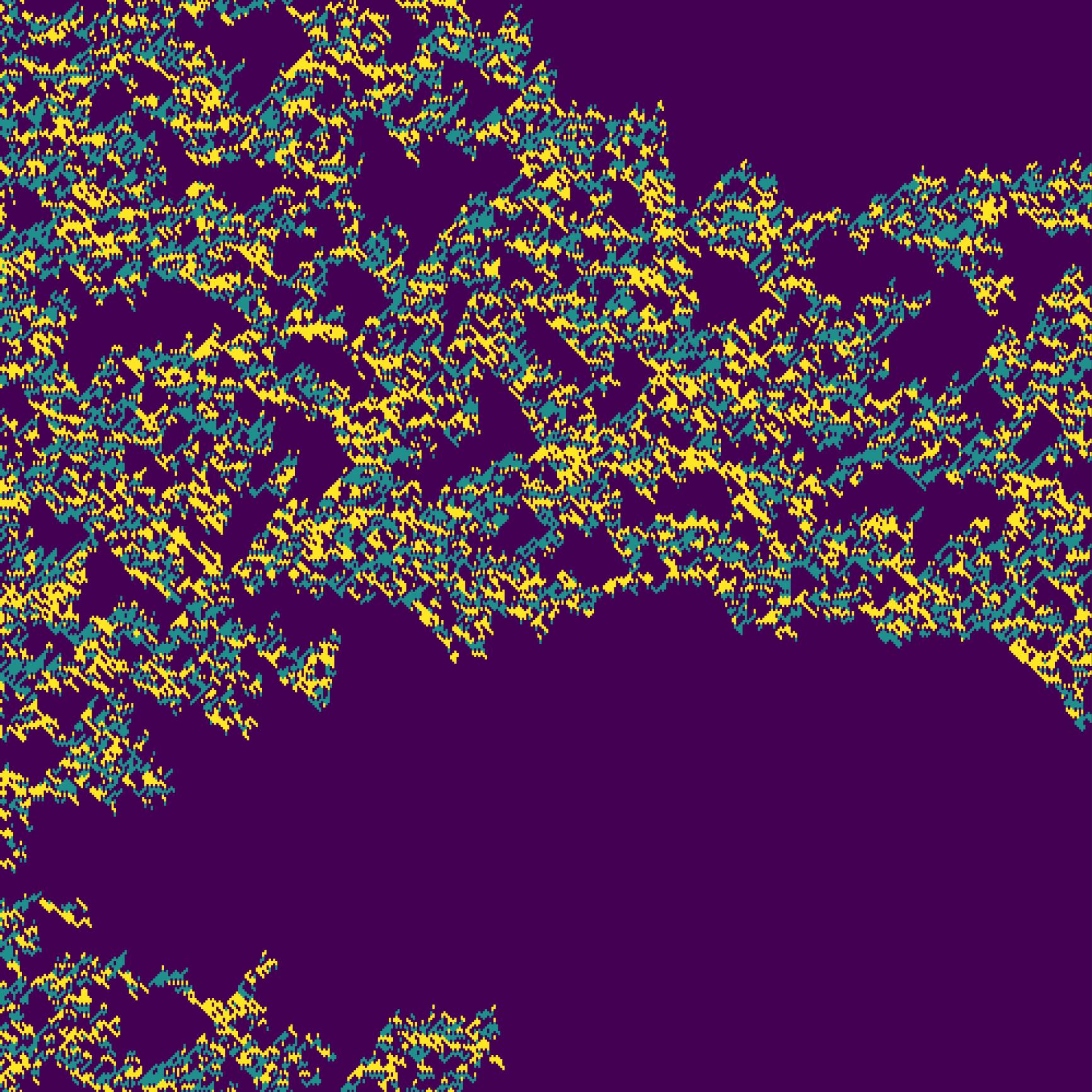}
    \caption{}
    \label{fig:1d}
  \end{subfigure}%
\caption{$D^+$-patterns: (a,c) MC data ($N=500$, $T=500$) for two configurations of the percolating phase at $q=0.9$; (a) $p=0.34$, (c) $p=0.31$; critical point of the transition into $D^+$ phase $p_c(q) \approx 0.317$; ochre/purple dots correspond to the filled/empty sites of the original lattice. Zoomed fragments of raw data
shown in insets. (b,d) The connected dipole (nn+nnn) patterns constructed from the data shown in (a,c); (b) $p>p_c(q)$, dipole pattern is not percolative; (d) $p<p_c(q)$, the system is in the percolating $D^+$ phase; yellow/teal dots correspond to connected $\pm 1$ dipoles, respectively, residing on the sites of the dipole lattice; the dark purple background corresponds to the sites without dipoles or to disconnected dipoles without ancestors.}
\label{fig:dipole+_pattern_example}
\end{figure*}

The examples of the raw MC data on the both sides from the transition into $D^+$ phase are shown in Fig.~\ref{fig:dipole+_pattern_example}, along the extracted $D^+$ patterns. The latter confirm the transition with the (nn+nnn) dipole percolative pattern.

%%%%%%%%%%%%%%%%%%%%%%%%%%%%%%%%%%%%%%%%%%%%%%%%%%%%%%%%%%%%%%%%%%%%%%%%%%%
%%%%%%%%%%%%%%%%%%%%%%%%%%%%%%%%%%%%%%%%%%%%%%%%%%%%%%%%%%%%%%%%%%%%%%%%%%%
%%%%%%%%%%%%%%%%%%%%%%%%%%%%%%%%%%%%%%%%%%%%%%%%%%%%%%%%%%%%%%%%%%%%%%%%%%%

\begin{figure*}[h]
\begin{subfigure}{0.4\textwidth}
    \includegraphics[width=7.2cm]{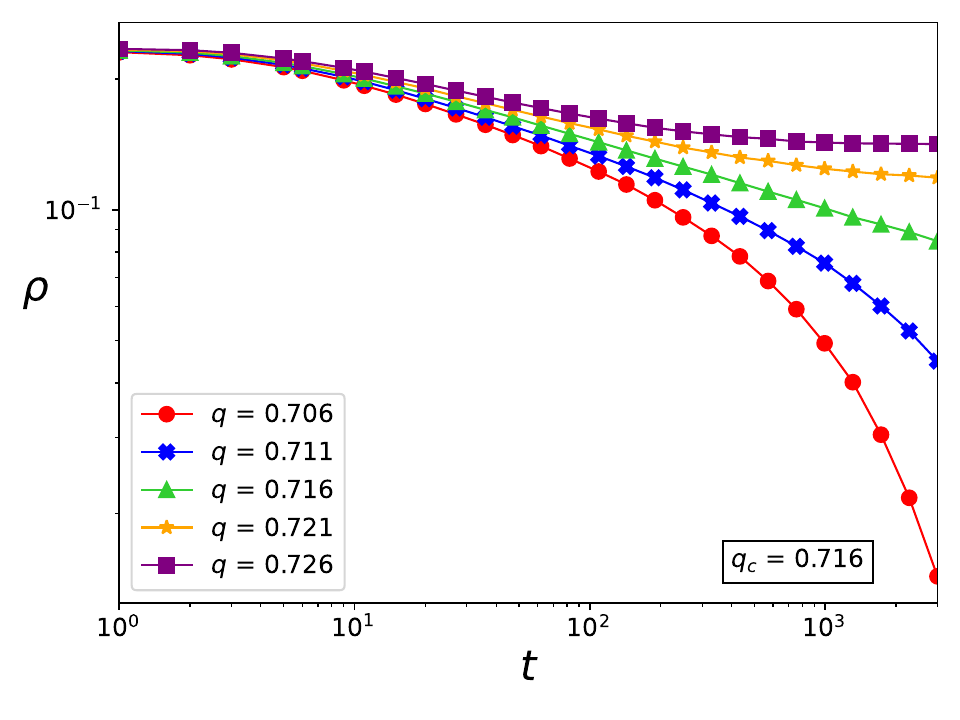}
    \caption{} \label{fig:1a}
  \end{subfigure}%
\begin{subfigure}{0.4\textwidth}
    \includegraphics[width=7.2cm]{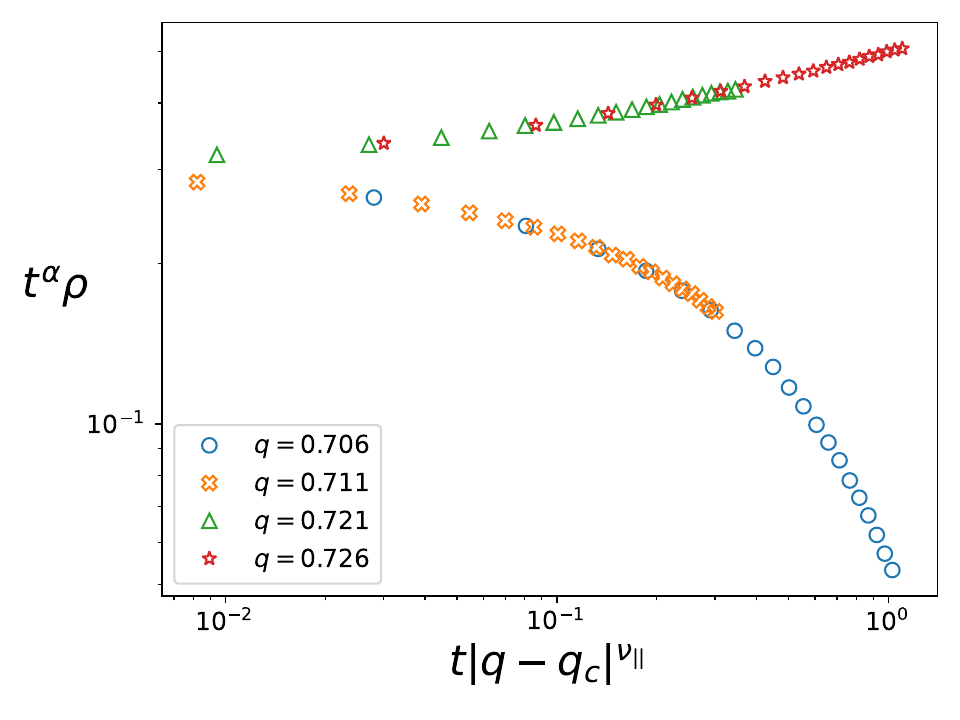}
    \caption{} \label{fig:1b}
  \end{subfigure}%

\begin{subfigure}{0.4\textwidth}
\hspace*{-7mm}
\includegraphics[width=6.7cm]
    {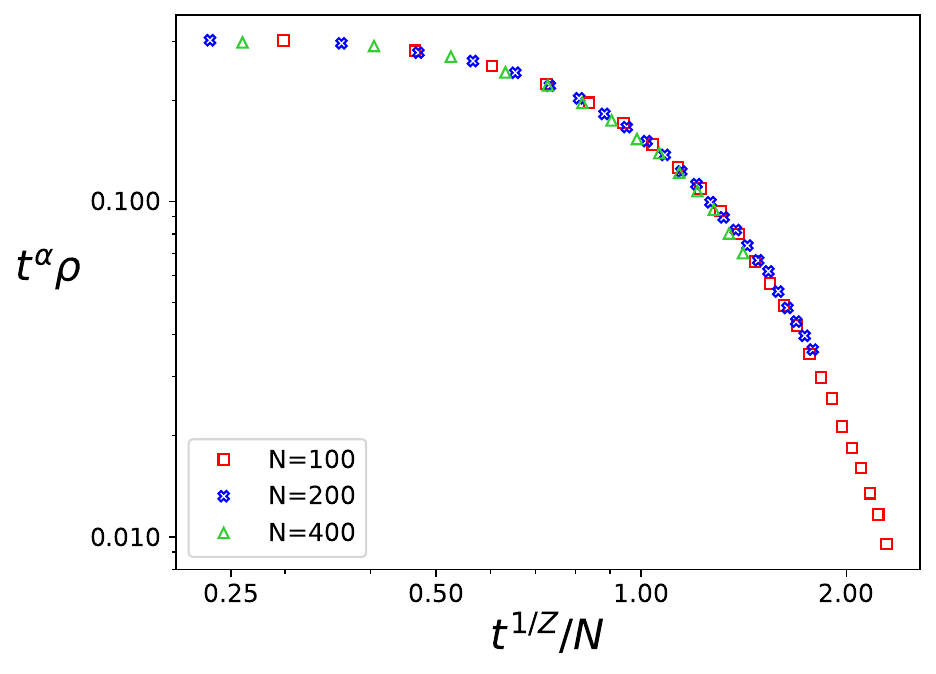}
    \caption{} \label{fig:1c}
  \end{subfigure}%
\begin{subfigure}{0.4\textwidth}
    \includegraphics[width=7.2cm]{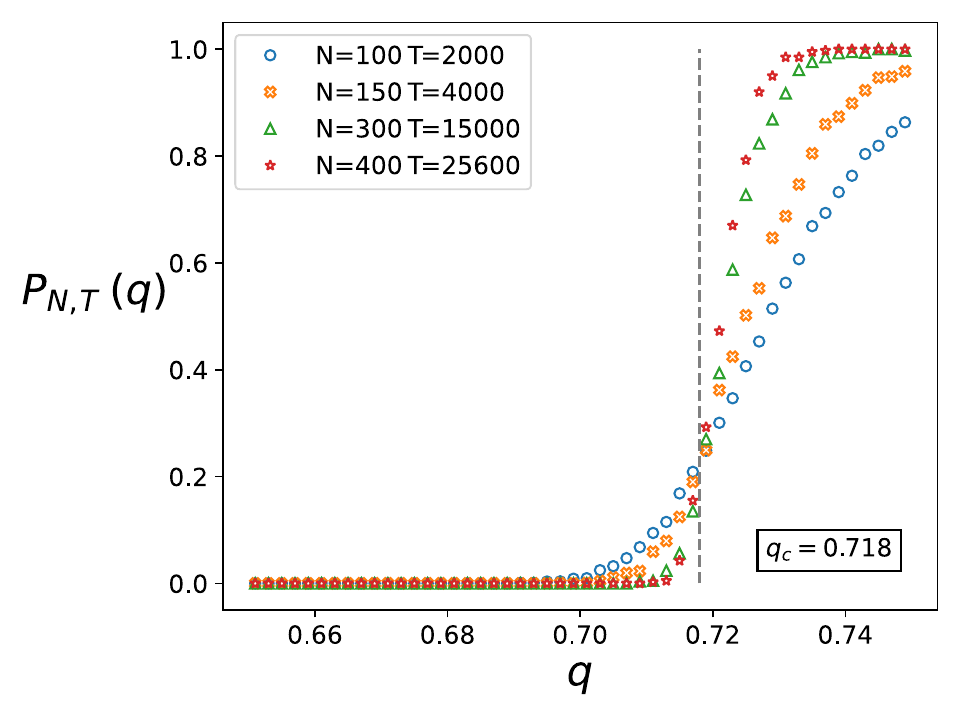}
    \caption{} \label{fig:1d}
  \end{subfigure}%
\caption{$D^+$ phase: (a) Density of the connected (nn+nnn) dipoles $\rho(t)$ for $p=0.2$ and series of $q$ near $q_c \approx 0.7158$. The system size is $N=3000$ and $T=100000$. The middle line corresponds to the critical slowing down $\rho(t) \propto t^{- \alpha}$ with $\alpha \approx 0.1590$.  (b) Collapse of the curves from (a) onto a single scaling function. Fitting gives the values of $q_c$ and $\nu_\parallel$. (c) Collapse of the order parameter relaxation $t^{\alpha}\rho(t)$ at the critical point  for different sizes ($N=100;~200;~400$) yields the critical index $z$.  (d) Fraction of connected dipoles $P_{N,T}(q)$ for different sizes $N \times T$, all curves intersect at  $q_c \approx 0.718$. The critical values of $q_c$ obtained from (a), (b) and (d) agree within $\sim 0.01 \% $  }
\label{fig:D+SC}
\end{figure*}

Some examples of the finite-size scaling analysis of the $Q^+$-phase are given in
Fig.~\ref{fig:D+SC}. The results for critical points $q_{c}$ and indices are collected in Table~\ref{Tab:D+tr}.

%%%%%%%%%%%%%%%%%%%%%%%%%%%%%%%%%%%%%%%%%%%%%%%%%%%%%%%%%%%%%%%%%%%%%%%%%%%%
\begin{table}[h]
\centering
\caption{Critical points $q_{c}$ and critical indices for the transition into $D^+$ phase for several values of $p$. Scaling analysis presented in Fig.~\ref{fig:D+SC} yields parameters shown in bold.}
{%
\begin{tabular}{|l|l|l|l|l|l|l|}
\hline
$p$ & $q_c$ & $\alpha$ & $\nu_{||}$ & z & $\beta = \alpha\nu_{||}$ & $\nu_{\perp} = \nu_{||} / z $ \\ \hline
0.0000 & 0.6441 & 0.1590 & 1.72 & 1.56 & 0.27(3) & 1.10(2) \\ \hline
\textbf{0.2000} & \textbf{0.7158} & \textbf{0.1590} & \textbf{1.72} & \textbf{1.56} & \textbf{0.27(3)} & \textbf{1.10(2)} \\ \hline
0.3000 & 0.8170 & 0.1610 & 1.72 & 1.55 & 0.27(7) & 1.10(9) \\ \hline
0.3270 & 1.0000 & 0.1590 & 1.72 & 1.58 & 0.27(3) & 1.08(9) \\ \hline
\end{tabular}%
}
\label{Tab:D+tr}
\end{table}
%%%%%%%%%%%%%%%%%%%%%%%%%%%%%%%%%%%%%%%%%%%%%%%%%%%%%%%%%%%%%%%%%%%%%%%%%%%%

%
%
%
%%%%%%%%%%%%%%%%%%%%%%%%%%%%%%%%%%%%%%%%%%%%%%%%%%%%%%%%%%%%%%%%%%%%%%%%%%%%
%%%%%%%%%%%%%%%%%%%%%%%%%%%%%%%%%%%%%%%%%%%%%%%%%%%%%%%%%%%%%%%%%%%%%%%%%%%%
%%%%%%%%%%%%%%%%%%%%%%%%%%%%%%%%%%%%%%%%%%%%%%%%%%%%%%%%%%%%%%%%%%%%%%%%%%%%
%%%%%%%%%%%%%%%%%%%%%%%%%%%%%%%%%%%%%%%%%%%%%%%%%%%%%%%%%%%%%%%%%%%%%%%%%%%%
%
%
%

\begin{figure*}[h]
\begin{subfigure}{0.4\textwidth}
    \includegraphics[width=7.2cm]{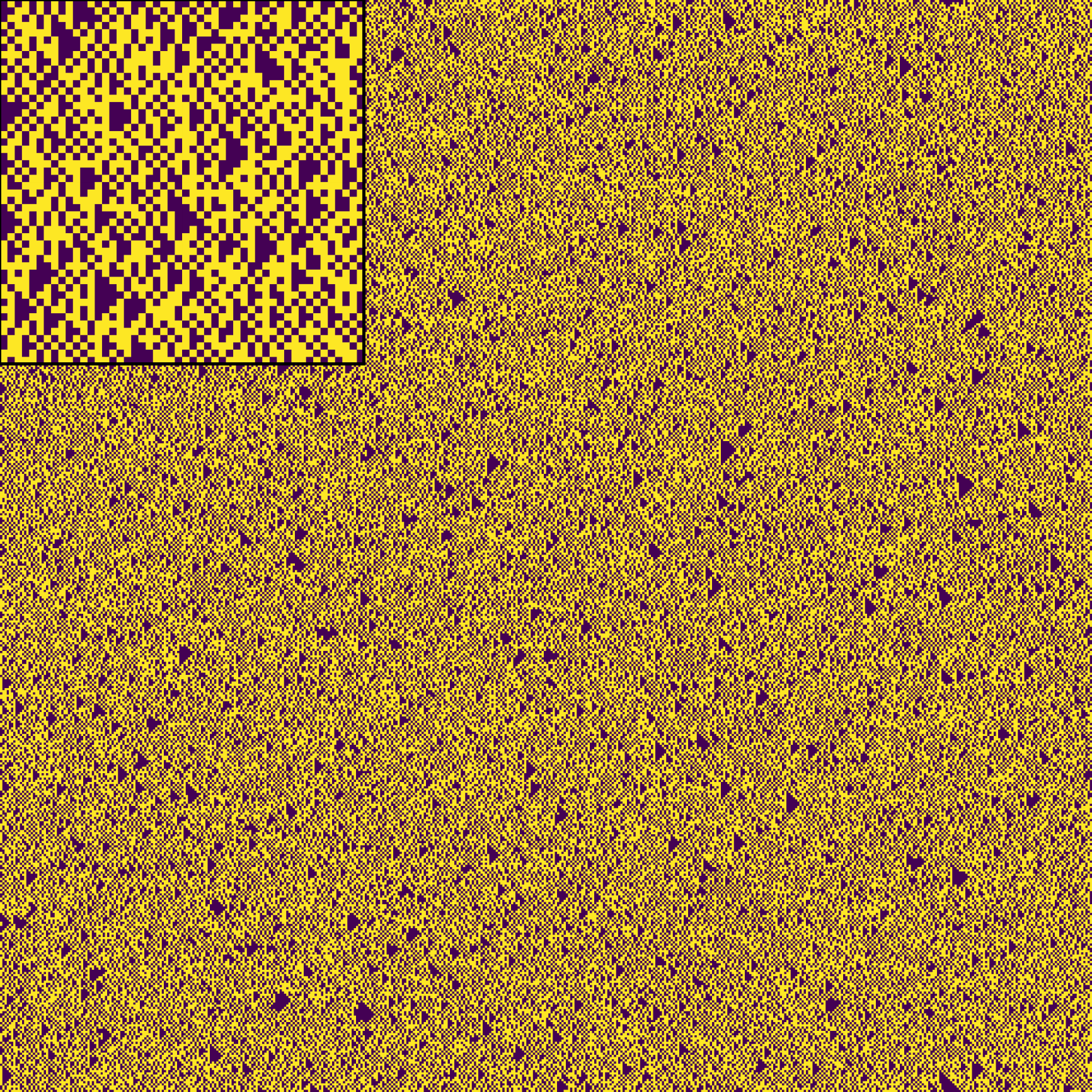}
    \caption{}
    \label{fig:1a}
  \end{subfigure}%
    \hspace{0.03\textwidth}
\begin{subfigure}{0.4\textwidth}
    \includegraphics[width=7.2cm]{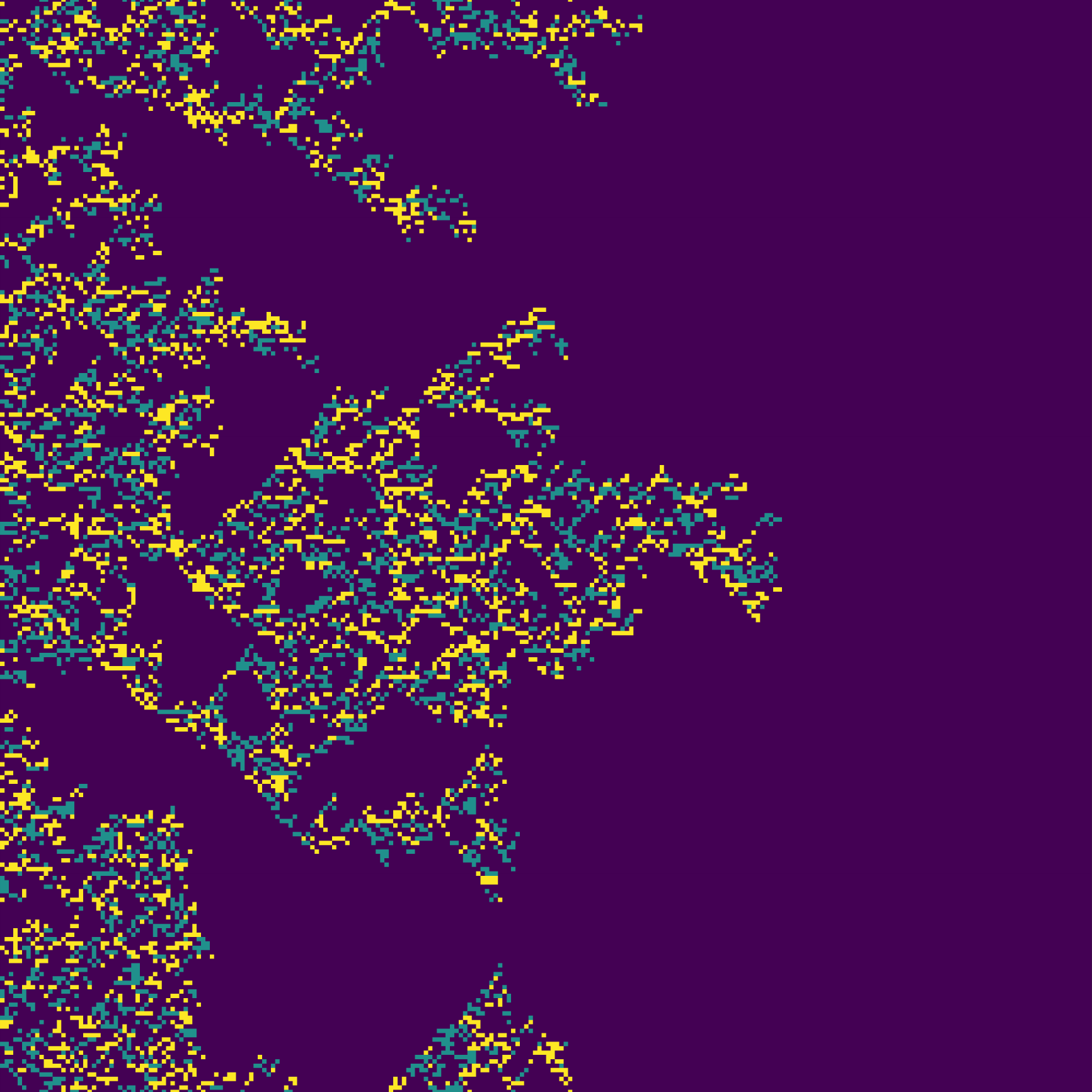}
    \caption{}
    \label{fig:1b}
  \end{subfigure}%

\begin{subfigure}{0.4\textwidth}
\includegraphics[width=7.2cm] {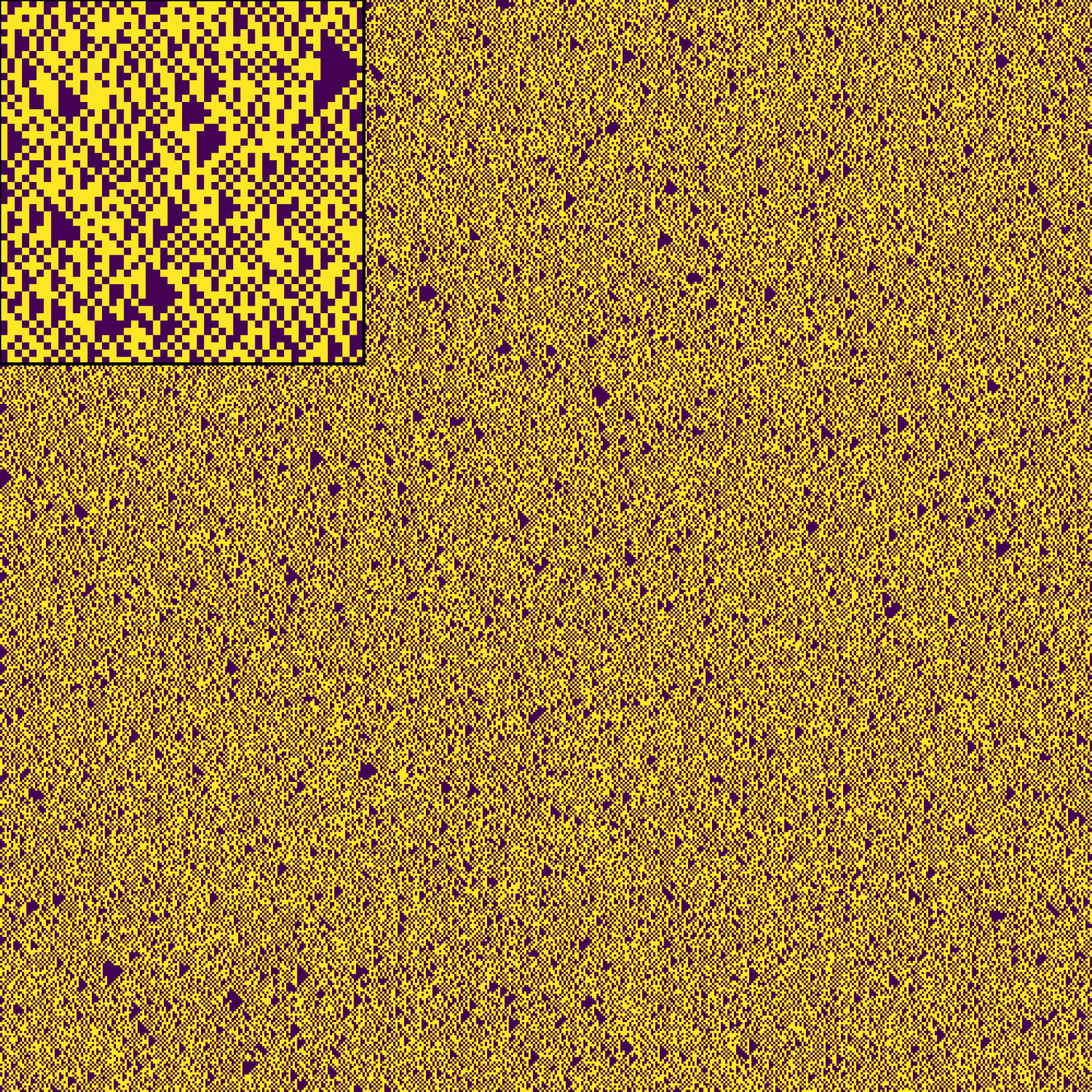}
    \caption{}
    \label{fig:1c}
  \end{subfigure}%
    \hspace{0.03\textwidth}
\begin{subfigure}{0.4\textwidth}
    \includegraphics[width=7.2cm]{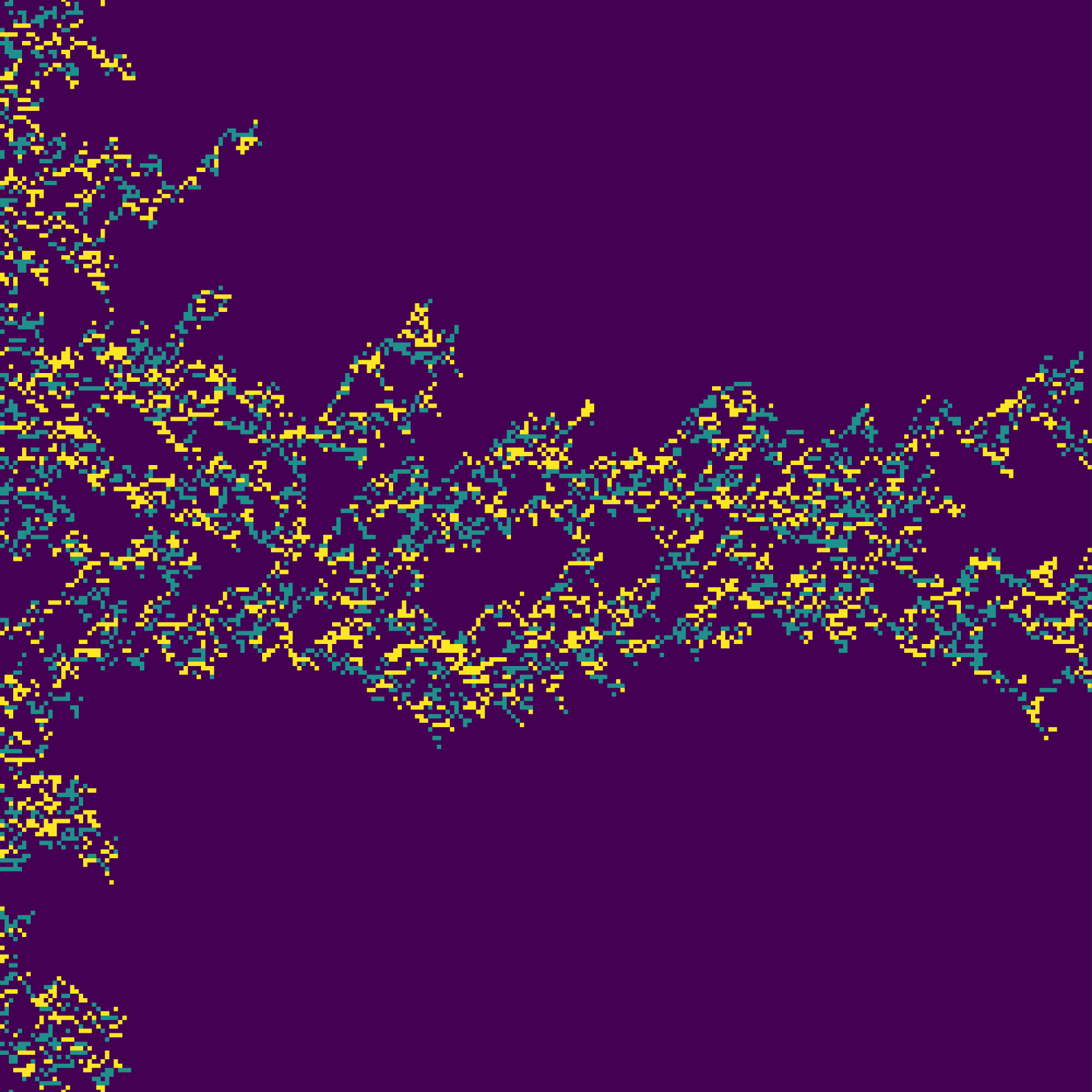}
    \caption{}
    \label{fig:1d}
  \end{subfigure}%
\caption{$Q^+$-patterns: (a,c) MC data ($N=500$, $T=500$) for two configurations of the percolating phase at $q=0.9$; (a) $p=0.35$, (c) $p=0.34$; critical point of the transition into $Q^+$ phase $p_c(q) \approx 0.337$; ochre/purple dots correspond to the filled/empty sites of the original lattice. Zoomed fragments of raw data shown in insets. (b,d) The connected  (nn+nnn) quadrupole  patterns constructed from the data shown in (a,c); (b) $p>p_c(q)$, $Q^+$-pattern is not percolative; (d) $p<p_c(q)$, the system is in the percolating $Q^+$ phase; yellow/teal dots correspond to connected $\pm 1$ quadrupoles, respectively, residing on the sites of the quadrupole lattice; the dark purple background corresponds to the sites without quadrupoles or to disconnected quadrupoles without ancestors. }
\label{fig:quadrupole+_pattern_example}
\end{figure*}

\begin{figure*}[h]
\begin{subfigure}{0.4\textwidth}
    \includegraphics[width=7.2cm]{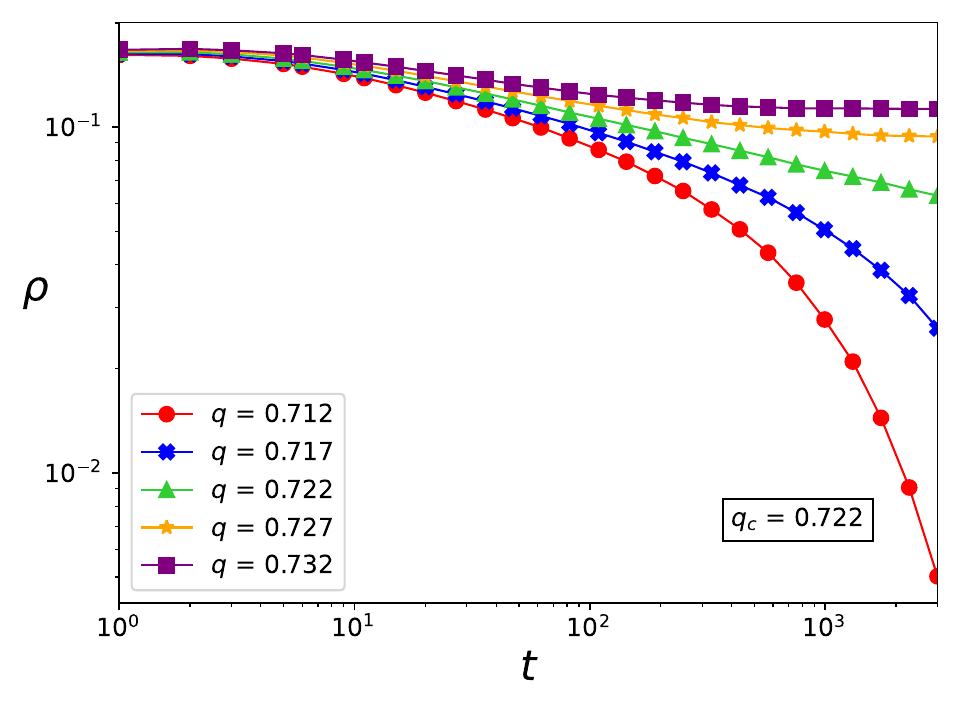}
    \caption{} \label{fig:3a}
  \end{subfigure}%
\begin{subfigure}{0.4\textwidth}
    \includegraphics[width=7.2cm]{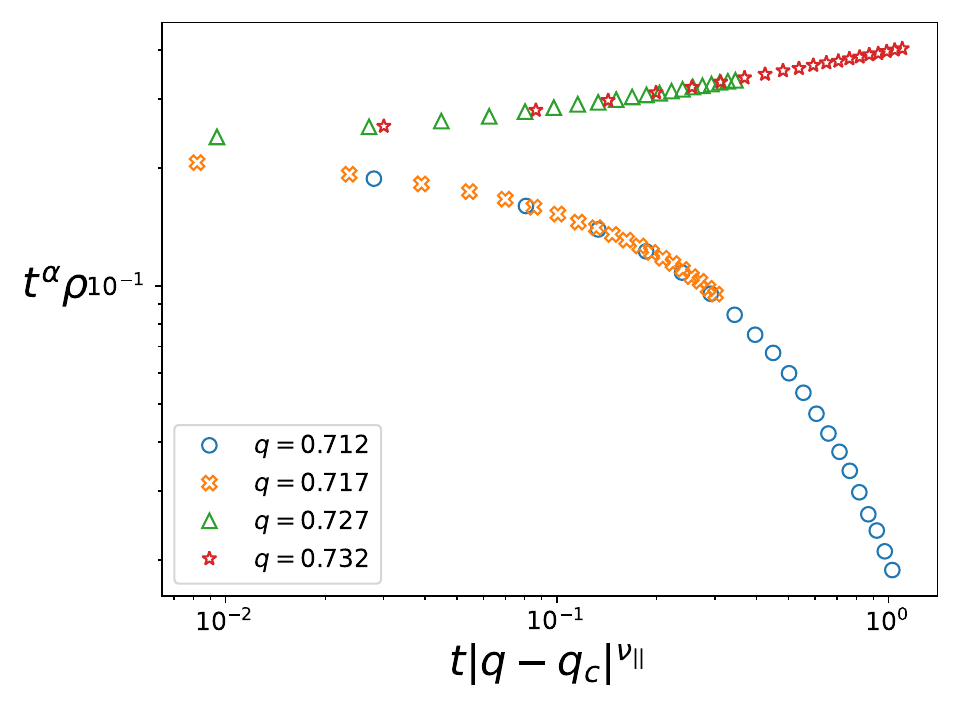}
    \caption{} \label{fig:3b}
  \end{subfigure}%

\begin{subfigure}{0.4\textwidth}
\hspace*{-7mm}
\includegraphics[width=6.7cm]
    {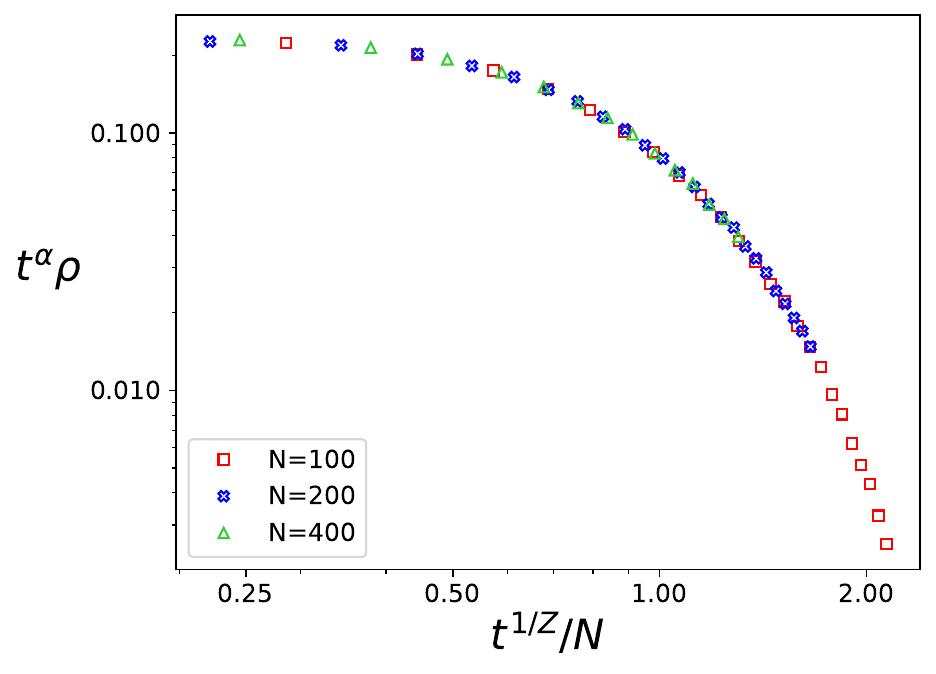}
    \caption{} \label{fig:3c}
  \end{subfigure}%
\begin{subfigure}{0.4\textwidth}
    \includegraphics[width=7.2cm]{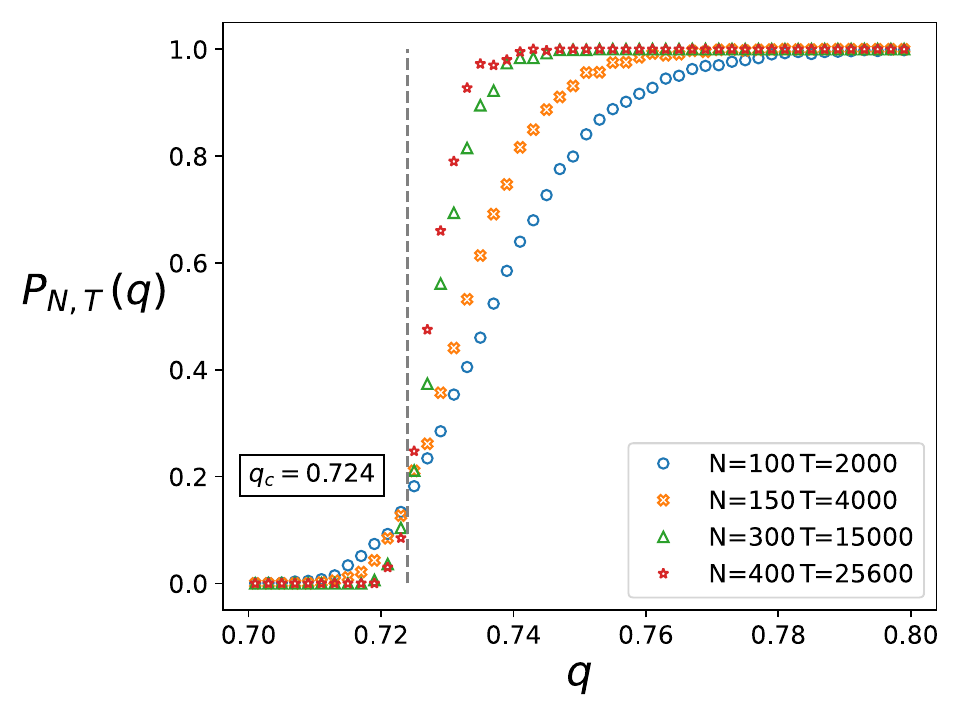}
    \caption{} \label{fig:3d}
  \end{subfigure}%
\caption{ $Q^+$-phase: (a) Density of the connected (nn+nnn) quadrupoles $\rho(t)$ for $p=0.2$ and series of $q$ near $q_c \approx  0.7219$. The system size is $N=3000$ and $T=100000$. The middle line corresponds to the critical slowing down $\rho(t) \propto t^{- \alpha}$ with $\alpha \approx 0.1595$.  (b) Collapse of the curves from (a) onto a single scaling function. Fitting gives the values of $q_c$ and $\nu_\parallel$. (c) Collapse of the order parameter relaxation $t^{\alpha}\rho(t)$ at the critical point  for different sizes
 ($N=100;~200;~400$) yields the critical index $z$. (d) Fraction of connected quadrupoles $P_{N,T}(q)$ for different sizes $N \times T$, all curves intersect at  $q_c \approx 0.724$. The critical values of $q_c$ obtained from (a), (b) and (d) agree within $\sim 0.01 \% $}
\label{Q+SC}
\end{figure*}

To demonstrate the occurrence of the (nn+nnn) quadruple percolative pattern,
two examples of the raw MC data on the both sides from the transition into $Q^+$ phase are shown in Fig.~\ref{fig:quadrupole+_pattern_example}, along the extracted patterns.
Examples of the finite-size scaling analysis of the $Q^+$-phase are given in
Fig.~\ref{Q+SC}, along with several critical points $q_{c}$ and indices in Table~\ref{Tab:Q+tr}.

%
%
%%%%%%%%%%%%%%%%%%%%%%%%%%%%%%%%%%%%%%%%%%%%%%%%%%%%%%%%%%%%%%%%%%%%%%%%%%%%
%%%%%%%%%%%%%%%%%%%%%%%%%%%%%%%%%%%%%%%%%%%%%%%%%%%%%%%%%%%%%%%%%%%%%%%%%%%%
%%%%%%%%%%%%%%%%%%%%%%%%%%%%%%%%%%%%%%%%%%%%%%%%%%%%%%%%%%%%%%%%%%%%%%%%%%%%
%%%%%%%%%%%%%%%%%%%%%%%%%%%%%%%%%%%%%%%%%%%%%%%%%%%%%%%%%%%%%%%%%%%%%%%%%%%%
%
%
%

%
%%%%%%%%%%%%%%%%%%%%%%%%%%%%%%%%%%%%%%%%%%%%%%%%%%%%%%%%%%%%%%%%%%%%%%%%%%%%
%%%%%%%%%%%%%%%%%%%%%%%%%%%%%%%%%%%%%%%%%%%%%%%%%%%%%%%%%%%%%%%%%%%%%%%%%%%%
\begin{table}[h]
\centering
\caption{Critical points $q_{c}$ and critical indices for the transition into $Q^+$ phase for several values of $p$. Scaling analysis presented in Fig.~\ref{Q+SC} yields parameters shown in bold.}
{%
\begin{tabular}{|l|l|l|l|l|l|l|}
\hline
$p$ & $q_c$ & $\alpha$ & $\nu_{||}$ & z & $\beta = \alpha\nu_{||}$ & $\nu_{\perp} = \nu_{||} / z $ \\ \hline
0.0000 & 0.6472 & 0.1595 & 1.72 & 1.54 & 0.27(4) & 1.11(6) \\ \hline
0.1000 & 0.6807 & 0.1590 & 1.72 & 1.58 & 0.27(3) & 1.08(8) \\ \hline
\textbf{0.2000} & \textbf{0.7219} & \textbf{0.1595} & \textbf{1.72} & \textbf{1.55} & \textbf{0.27(4)} & \textbf{1.10(9)} \\ \hline
0.3000 & 0.8167 & 0.1595 & 1.72 & 1.55 & 0.27(4) & 1.10(9) \\ \hline
0.3500 & 1.0000 & 0.1595 & 1.72 & 1.56 & 0.27(4) & 1.10(2) \\ \hline
\end{tabular}%
}
\label{Tab:Q+tr}
\end{table}
%%%%%%%%%%%%%%%%%%%%%%%%%%%%%%%%%%%%%%%%%%%%%%%%%%%%%%%%%%%%%%%%%%%%%%%%%%%%
%

%
%
%
%%%%%%%%%%%%%%%%%%%%%%%%%%%%%%%%%%%%%%%%%%%%%%%
%%%%%%%%%%%%%%%%%%%%%%%%%%%%%%%%%%%%%%%%%%%%%%%%%%%%%%%%%%%%%%%%%%%%%%%%%%%%
%%%%%%%%%%%%%%%%%%%%%%%%%%%%%%%%%%%%%%%%%%%%%%%%%%%%%%%%%%%%%%%%%%%%%%%%%%%%
%
%
%

%\end{widetext}
%
%
\end{appendix}
%%%%%%%%%%%%%%%%%%%%%%%%%%%%%%%%%%%%%%%%%%%%%%%%%%%%%%%%%%%%%%%%%%%%%%%%%%%%%%
%%%%%%%%%%%%%%%%%%%%%%%%%%%%%%%%%%%%%%%%%%%%%%%%%%%%%%%%%%%%%%%%%%%%%%%%%%%%%%
%%%%%%%%%%%%%%%%%%%%%%%%%%%%%%%%%%%%%%%%%%%%%%%%%%%%%%%%%%%%%%%%%%%%%%%%%%%%%%
%%%%%%%%%%%%%%%%%%%%%%%%%%%%%%%%%%%%%%%%%%%%%%%%%%%%%%%%%%%%%%%%%%%%%%%%%%%%%%
%
%
%

%
%
\end{document}